\documentclass[Afour,sagev,times,fleqn]{sagej}

\usepackage{moreverb,url}

\usepackage[colorlinks,bookmarksopen,bookmarksnumbered,citecolor=red,urlcolor=red]{hyperref}

\newcommand\BibTeX{{\rmfamily B\kern-.05em \textsc{i\kern-.025em b}\kern-.08em
T\kern-.1667em\lower.7ex\hbox{E}\kern-.125emX}}

\def\volumeyear{}

\usepackage{fleqn}
\usepackage{graphicx}
\usepackage{geometry}

\usepackage{epstopdf} 

\usepackage{tikz}
\usetikzlibrary{shapes.misc}

\usepackage{paralist}
\usepackage[inline]{enumitem}

\usepackage{amsmath,amssymb}

\usepackage[T1]{fontenc}
\usepackage[latin1]{inputenc} 

\usepackage{subcaption}

\usepackage{tabularx}

\usepackage{float}

\usepackage{xcolor}
\usepackage{soul}

\usepackage{paracol}
\usepackage{multicol}

\usepackage{verbatim}

\definecolor{gray97}{gray}{.97}
\definecolor{gray75}{gray}{.75}
\definecolor{gray45}{gray}{.45}

\newcommand{\evt}{\lozenge}
\newcommand{\A}{\mathcal{A}}
\newcommand{\E}{\mathcal{E}}
\newcommand{\nat}{\mathbb{N}}
\newcommand{\TA}{\mathit{TA}}
\renewcommand{\Re}{\mathbb{R}}

\newtheorem{definition}{Definition}

\newfloat{Pattern}{t}{patt}

\newif\ifcomments
\commentstrue 

\newcommand{\ariel}[1]{\ifcomments\sethlcolor{pink}\hl{A: #1}\fi}

\begin{document}

\runninghead{Gonzalez, Cristiá and Luna}

\title{Verification of Quantitative Temporal Properties in RealTime-DEVS}

\author{Ariel Gonz\'alez\affilnum{1} and Maximiliano Cristi\'a\affilnum{2} and Carlos Luna\affilnum{3}}

\affiliation{\affilnum{1} Universidad Nacional de Río Cuarto, Río Cuarto, Argentina\\
\affilnum{2} Universidad Nacional de Rosario and CIFASIS, Rosario, Argentina\\
\affilnum{3} Universidad de la Rep\'ublica, Montevideo, Uruguay}

\begin{abstract}

Real-Time DEVS (RT-DEVS) can model systems with quantitative temporal requirements. Ensuring that such models verify that kind of temporal properties requires to use something beyond simulation. In this work we use the model checker Uppaal to verify a class of recurrent quantitative temporal properties appearing in RT-DEVS models,
even though Uppaal cannot deal in general with this kind of properties. In order to overcome these limitations we use the technique known as automata observer. Secondly, by introducing mutations to quantitative temporal properties we are able to find errors in RT-DEVS models and their implementations. A case study from the railway domain is presented.
\end{abstract}
 

\maketitle

\section{Introduction}
\label{Introduccion}

Critical systems, in particular those containing timing constraints, must be thoroughly verified. These systems must produce not only the right answers but must produce them in the right moment, no sooner no later. The timely reaction is as important as the reaction itself. These systems are called \emph{real-time systems}. In some cases, the failure to accomplish a timing constraint has minor consequences---these are called \emph{soft timing constraints}. For instance, in the domotics domain if the real-time requirement ``lights must be turned off after 10 seconds of closing the door'' is met a few seconds later, the failure is not severe. In other application domains, the consequences of producing the right answer but in the wrong time frame can be catastrophic---these are called \emph{hard timing constraints}. For instance, if a car braking system reacts too late due to a software error it may cause a deadly accident; likewise, an intensive care unit device reacting too late can cause severe damages to the health of a patient. This work contributes to the verification of this kind of systems.

In general, the modeling and simulation (M\&S) community checks their models by running some simulations. Simulations are adequate to understand the behavior of a model \cite{cui2018pulsim,chirkin2017,zeigler2000theory}, but they fall short when it comes down to ensure the correctness of a model with respect to a property. In order to ensure that a model does not violate a given property we should run all possible simulations---which in general is unfeasible. On the other hand, tools such as model checkers \cite{Clarke2012} can \emph{automatically prove} that a model verifies a given property. However, model checkers may face the state explosion problem or lack the expressiveness necessary for some kind of systems. Hence, M\&S and model checking should be combined to get the best of both worlds. In this work we propose a technique combining the Real-Time DEVS (RT-DEVS) formalism \cite{RTDEVS} with the Uppaal model checker \cite{Uppaal04} to verify certain classes of real-time properties.

There are several languages that can be used to formally model or specify hard real-time systems---temporal Petri Nets \cite{TemporalPetriNets}, Discrete Event System Specification (DEVS) \cite{devsbook}, Timed Automata \cite{AlurDill94}, etc. Furthermore, there exists software tools assisting in the specification and verification of the models written in those languages. RT-DEVS is a variant of classic DEVS which foster the specification of real-time systems. In RT-DEVS the time advance function returns a time interval because the occurrence of an event is better represented by an interval. This is a recurrent feature of real-time systems. That is, even in hard real-time systems answers are expected to occur within a time interval rather than in a time instant \cite{DBLP:books/aw/Lamport2002}. This is so because, ultimately, the hardware might not be able to produce a certain answer in a specific time instant because it may be busy receiving inputs or producing other outputs. Timing constraints expressed as events occurring in precise time intervals are called \emph{quantitative temporal requirements} (QTR) or \emph{quantitative temporal properties} (QTP).

On the other hand, there are several languages that can be used to formally specify the temporal properties that models (or systems) must verify. Logic languages such as \emph{Linear Temporal Logic} (LTL) \cite{DBLP:conf/focs/Pnueli77} and \emph{Computation Tree Logic } (CTL)  \cite{DBLP:conf/focs/Pnueli77,CTL} are used to describe real-time properties. One of the main advantages of these logics is that they are supported by state-of-the-art model checkers \cite{nuSMV,Holzmann97}. 
However, these logics do not allow for the specification of QTP. Conversely, the \emph{Timed Computational Tree Logic} (TCTL) \cite{TCTL} and the \emph{Metric Temporal Logic} (MTL) \cite{Koymans1990,Ouaknine2007} are designed for the specification of QTP although, at best, they are partially supported by model checkers.

Therefore, in this work we propose a technique for the verification of some classes of QTR expressed in RT-DEVS and MTL by using the Uppaal model checker, which partially supports TCTL. 
Given that Uppaal does not directly admit the analysis of QTR, we propose a process where a RT-DEVS model is transformed into an automaton which is controlled by an automaton observer \cite{MGT2009,BackesWGK16}. Furthermore, we show how Uppaal can be used to generate test cases to test temporal properties on the implementation of RT-DEVS models.

\subsection{Contributions}
\label{Contribucion}
\begin{figure}
	\begin{center}
		\includegraphics [scale=0.27]{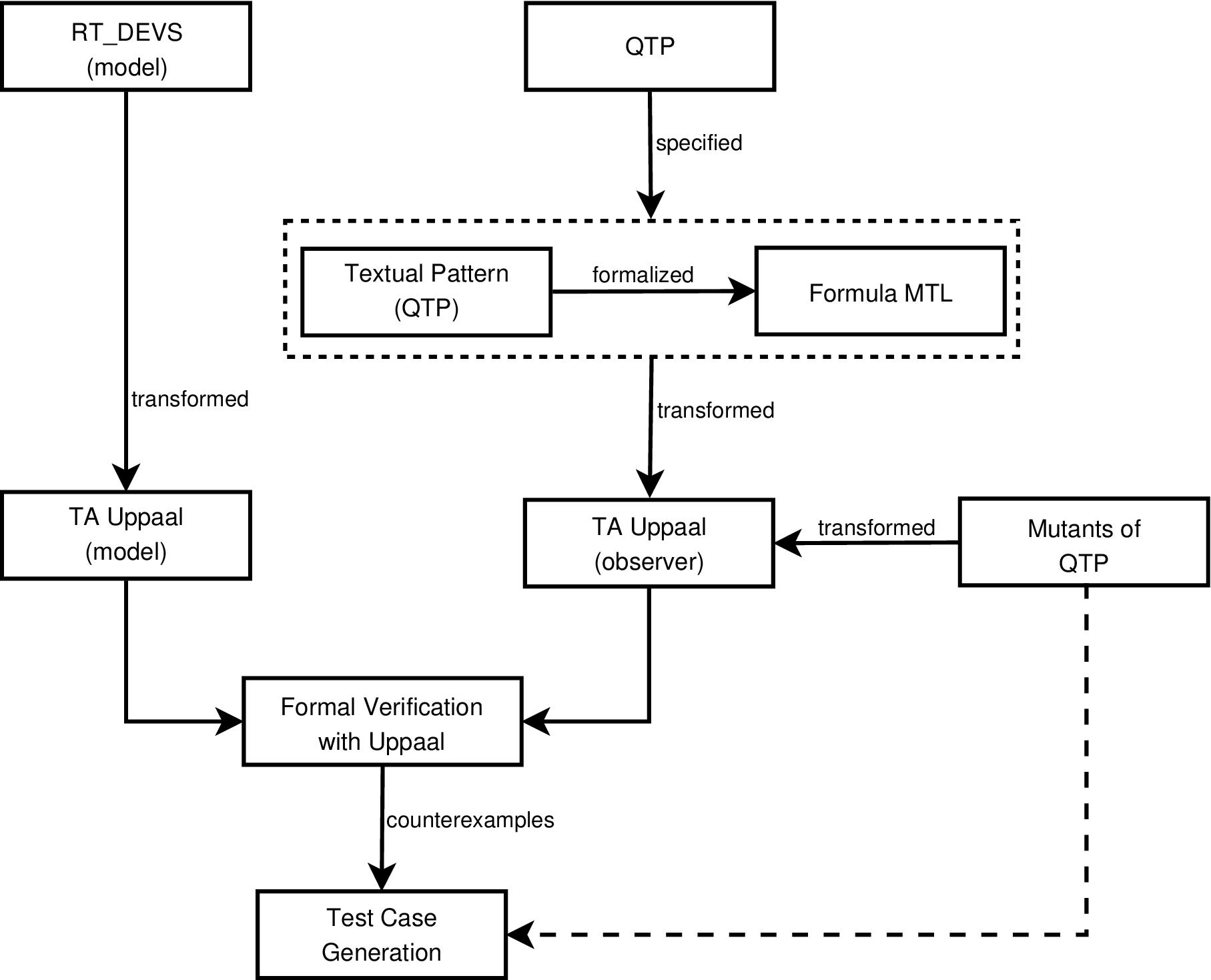}
	\end{center}
	\caption{Activity Flow of the Formal Verification Process}
	\label{fig:WorkflowProceso} 
\end{figure}

The contributions of the present work can be better understood by looking at Fig. \ref{fig:WorkflowProceso}. Engineers write a RT-DEVS model which must verify some QTP. We propose to transform both into Timed Automata (TA) \cite{Alur1992,AlurDill94} that are fed into the Uppaal model checker, which can automatically prove whether or not the model verifies the QTP. However, since QTP may be hard to formalize a set of patterns of temporal formulas is provided (dotted-line box). In this way the engineer can start with an informal, familiar description of the temporal property and then formalize it with the help of the pattern documentation. Finally, mutants of QTP and counterexamples generated by Uppaal help engineers to find temporal errors in their models, that can also be used to generate test cases for an implementation of the model.

Hence, our contributions are the following:
\begin{itemize}
\item The main contribution is a technique for the verification of QTP expressed by means of RT-DEVS and MTL using the Uppaal model checker.
We are not aware of methods departing from RT-DEVS models featuring QTP that provide insights on how these timing properties can be formally verified with a model checker.
In particular, our approach makes use of \emph{automata observer} \cite{MGT2009,BackesWGK16} and some powerful Uppaal elements. The patterns of temporal formulas are borrowed from previous work by the authors \cite{cibseGCL19}. 
\item When it comes to real-time systems, engineers are faced with the daunting task of verifying complex behaviors including intricate real-time constraints.
Hence, a second contribution of this paper is a technique that uses  mutants of temporal properties and Uppaal to help engineers to find temporal errors in RT-DEVS models. This technique is complementary to a simulation-based verification process. Although mutants of temporal formulas have also been proposed by the authors in their earlier works, here mutants are implemented in Uppaal thus integrating them with the rest of the method.
\item A third contribution is the extension of the above technique to generate test cases for the implementation of the RT-DEVS models.
\end{itemize}

The contributions are illustrated by means of a case study from the railway domain. In particular the case study includes the analysis of widely used QTP such as \emph{Time-Bounded Response}, \emph{Time-Restricted Precedence} and \emph{Conditional Security}.

\subsection{Structure of the paper}
\label{organizacion}

Section \ref{Trabajos Relacionados} presents and analyze the literature concerned with the formal verification of DEVS models and some of its variants. 
In Section \ref{Nociones Preliminares} we introduce some background about RT-DEVS, Timed Automata, MTL and model checking.
Then, in Section \ref{Caso de Estudio: Train-Gate} the case study used as a running example across the paper to validate our proposal is presented.
Section \ref{Metodologia de Verificacion de RT-DEVS} introduces the core of our work, that is, the technique to verify with Uppaal some expressive and recurrent classes of QTP described with RT-DEVS and MTL. A key aspect of our technique is that users can use patterns of QTP to simplify the specification of complex temporal properties (Section \ref{Implementación de los Patrones de propiedades temporales cuantitativas}).
In Section \ref{Encontrado errores con mutantes de patrones temporales} we explain how ideas borrowed from mutation testing and specification mutation can be used to find timing errors in RT-DEVS models and their implementations.
Indeed, we show how mutations introduced to patters of QTP can help in debugging models and implementations.
Finally, our conclusions and possible future works are described in Section \ref{Conclusiones y Trabajos Futuros}. 
Appendix \ref{RCS RT-DEVS} presents a formal specification of the case study, whereas Appendix \ref{Patrones de propiedades temporales cuantitativas en Uppaal}  introduces some patterns of QTP not shown in the main text.

\section{Related Work}
\label{Trabajos Relacionados}

In the context of DEVS verification there are several works describing how to check properties. 
Song and Kim \cite{SK2005} propose a mechanism for the verification of some safety properties over RT-DEVS models. Properties are given in a logic similar to LTL. Communication between the RT-DEVS models is performed by means of a so-called \emph{clock matrix}. The authors do not use model checkers, but define an algorithm based on this matrix that builds a timed reachability tree that is used to analyze safety properties.
Furfano and Nigro \cite{FN2008} describe how RT-DEVS models can be translated into the TA supported by Uppaal. They also use ActorDEVS \cite{ActorDevs} to run simulations and Uppaal to verify some temporal properties.

Classic DEVS models can be translated into TA by means of the translation defined in terms of a \emph{simulation relation} by Dacharry and Giambiasi \cite{DG2007}. The simulation relation links the behavior of a DEVS model with that of the corresponding TA. The intention of the authors is to describe high-level properties with TA and low-level properties with DEVS. The work does not give clues about using the method for property verification.
Inostrosa-Psijas et al. \cite{PSIJAS2019} propose another translation between classic DEVS and TA by means of bisimulation. This translation is applied to a model of Apache Storm \cite{ApacheStorn}. The resulting TA are verified using Uppaal. 
Rational-TimeAdvance DEVS (RTA-DEVS) \cite{SaadawiW13,SWM12} are characterized by a time advance function accepting rational constants only. Saadawi and Wainer  propose a translation from RTA-DEVS into TA that are later verified with Uppaal. In the second paper the authors use their results to verify properties such as deadlock on the controller of an E-Puck robot. 
Gholami and Sarjoughian \cite{GS2017} define their own DEVS flavor by considering a finite number of states, inputs, outputs and internal and external transitions within some finite time interval. The authors develop their own Java tool to verify their models. Models and properties are both written in Java. 
Gonz\'alez, Cristi\'a and Luna propose a technique to find errors in real-time systems modeled in classic DEVS \cite{cibseGCL19}. Temporal properties are expressed in MTL. Since temporal properties, may be hard to formalize a set of patterns of temporal formulas is proposed. These patterns cover a wide class of recurrent temporal requirements. The formulas so generated are modified by applying a predefined set of \emph{mutations} (cf. to mutation testing \cite{Buchler11,TanSL04,Trakhtenbrot17} and specification mutation \cite{DBLP:conf/issta/StocksC93}). Later, simulations distinguishing the original property from a mutant are generated. If the model reacts as expected by the mutant property, the model does not verify the property. Decidability issues concerning MTL and, consequently, the lack of model checkers supporting it motivated a second work \cite{SCCC2021}. There, real-time systems are modeled by means of Timed Automata (TA) \cite{AlurDill94}.

As can be seen, the mentioned works translate some DEVS variants into the TA supported by Uppaal and use it to analyze the validity of some properties. Given that Uppaal implements a TCTL subset not including timed temporal modalities as those defined by Alur \cite{Alur1992}, none of these methods can verify QTP. For instance, none of these works can express properties such as \emph{P enables Q for k time-units}. Saadawi and Wainer show how one specific QTP---namely \emph{P must be followed by Q within k time-units}---could be checked by adding a Boolean variable and a clock.

Precisely, in the present work, we show how several key and recurrent QTP patterns appearing in RT-DEVS models can be verified with Uppaal.

\section{Background}
\label{Nociones Preliminares}

In this section we briefly recall the main notions used in the rest of the work: RT-DEVS (\ref{Formalismo RT-DEVS}), Timed Automata (\ref{Timed Automata}), Metric Temporal Logic (\ref{Metric Temporal Logic (MTL)}), and basic concepts on model checking (\ref{Model Checking}). These notions are the theoretical foundations for the proposed technique. Readers familiar with any of these can skip the corresponding section.

\subsection{Real-Time DEVS}
\label{Formalismo RT-DEVS}

RT-DEVS \cite{RTDEVS,SK2005} is a variant of the DEVS formalism better suited to model real-time systems. The main difference between RT-DEVS and DEVS is that the former allows the specification of the occurrence of an event within a time interval and not necessarily in a specific time instant.
\begin{definition}\label{def:rt-devs}
An atomic RT-DEVS model is a tuple $\left\langle X,Y,S,ta,\delta_{ext},\delta_{int},\lambda,\A,\omega,ti\right\rangle$, where:
\begin{enumerate*}[label=(\emph{\roman*})]
\item $X$  is the set of input events;
\item $Y=$ is the set of output events;
\item $S$ is the set of states;
\item $ta : S \rightarrow \mathbb{R}^{+}_{0,\infty} $ is the time advance function;
\item $\delta_{int}: S \rightarrow S$ is the internal transition function;
\item $\lambda: S \rightarrow Y$ is the output function;

%
\item $\A = \{a \mid t(a)\in \mathbb{R}^{+}_{0,\infty} \land a \notin \{X?,Y!,S{=}\}\}$, is the set of activities where $t(a)$ is the execution time of activity $a$, $X?$ is the action of receiving data from $X$, $Y!$ is the action of sending data through $Y$, and $S{=}$ is the action of modifying a state in $S$;
\item $\omega : S \rightarrow \A$ is the activity mapping function;
\item $ti : S \rightarrow \mathbb{R}^{+}_{0,\infty} \times \mathbb{R}^{+}_{0,\infty}$ is a time interval function such that if $ti(s)=[x, y]$ then $x \leq ta(s) \leq y$ and $x \leq t(a) \leq y$, with $a=\omega(s)$.
\item $\delta_{ext}: Q \times X \rightarrow S$ is the external transition function, with $Q = \{(s,e) | s\in S \land (ti(s)=[x, y] \implies 0 \leq e \leq y)\}$.
\end{enumerate*}
\end{definition}

As can be seen, an activity consumes time while the system is in a given state---$\omega$ returns the activity for each state. Transmitting a message, carrying out a task, etc. are examples of activities. 

In this paper we will use a simplified version of RT-DEVS where $ta(s)=t(a)$, with $a=\omega(s)$, for every state $s$. In this context the activity mapping function ($\omega$) can be ignored. Therefore, in this paper a RT-DEVS model is characterized as $\langle X,Y,S,\delta_{ext},\delta_{int},\lambda,ti\rangle$ . Then, our RT-DEVS models are DEVS models where the time advance function is replaced by the time interval function.

As with DEVS models, RT-DEVS models can be coupled by means of ports to build complex models. Given a RT-DEVS coupled model, it is possible to obtain an equivalent atomic RT-DEVS model, as in DEVS. 
In this case $X$ (inputs) and $Y$ (outputs) have the following form:
\begin{enumerate*}[label=(\emph{\roman*})]
\item $X = \{ (p,v) \mid p \in \mathit{InPorts} \land v \in X_p\}$ where $\mathit{InPorts}$ is the set of input ports and $X_p$ is the set of input values that can be received at port $p$;
\item $Y=\{(p,v) \mid p \in \mathit{OutPorts} \land v \in Y_p\}$ where $\mathit{OutPorts}$ is the set of output ports and $Y_p$ is the set of values that can be sent through port $p$.
\end{enumerate*}
When RT-DEVS models are coupled, input ports of one component are connected to output ports of other components, and vice versa \cite{Waigner2018}.

\subsection{Timed Automata}
\label{Timed Automata}

A Timed Automata (TA) \cite{AlurDill94, Bengtsson2004} is a finite automata extended with variables known as \emph{clocks} used to specify timed states and transitions. 

\begin{definition}
A Timed Automata is given by a tuple $\langle \Sigma,N,N_0,C,E,I \rangle$ where:
\begin{enumerate*}[label=(\emph{\roman*})]
\item $\Sigma$ is a finite set of actions;
\item $N$ is a finite set of locations, states or nodes; 
\item $N_0 \in N$ is the initial node;
\item $C$ is the finite set of clock variables, or just clocks;
\item $E \subseteq N \times \beta (C) \times \Sigma \times 2^C \times N$ is a set of transitions; and
\item $I : N \rightarrow \beta (C)$ is a function assigning clock constraints to the nodes.
\end{enumerate*}
In turn, $2^C$ is the powerset of $C$; and $\beta(C)$ is a set of clock constraints.
\end{definition}

A clock constraint is a conjunction of atomic predicates of the form: 
$x \sim n$ and $x-y \sim n$ with $x, y \in C$, $\sim\, \in \{<, >, =, \leq,\geq \}$ and $n \in \nat$, where $\nat$ is the set of natural numbers. A clock constraint appearing in a transition (node) is called \emph{guard} (\emph{invariant}).
Guards must be satisfied for the transition to take place. Invariants restrict the time the automata can remain in a node.

An expression of the form $n \xrightarrow{g,a,r} n'$ with $(n, g, a, r, n') \in E$ is a transition from node $n$ to node $n'$ where $g$ is a clock restriction, $a$ an action and $r$ is a set of clocks to be set to zero.
The state of a TA is a pair of the form $(n,u)$ where $n$ is the current node and $u:C \rightarrow \Re$, called \emph{valuation}, is a function that gives the value of each clock in $n$.
A TA can change the state by the mere passing of time (called \emph{delay}) or by executing an action. A delay is symbolized as $(n, u) \xrightarrow{d} (n, u + d)$, with $d \in \Re$. This means that the TA moves to a state with the same node but where clocks have been updated by the expression $u+d = \{(c,y) \mid c \in C \land y = u(c) + d\}$. The execution of an action is symbolized as $(n, u) \xrightarrow{a} (n' , u')$ where action $a$ fires a transition of the form $n \xrightarrow{g,a,r} n'$ such that: $u$ satisfies $g$; clocks in $r$ are set to zero; $u'= \{(c,y) \mid c \in C \land \textbf{ if } c \in r \textbf{ then } y = 0 \textbf{ else } y = u(c)\}$; and $u'$ satisfies $I(n')$.

\subsection{Metric Temporal Logic}
\label{Metric Temporal Logic (MTL)}

MTL is an extension of LTL. LTL is a popular formalism for the specification and verification of reactive concurrent systems. It allows the specification of a wide spectrum of temporal properties, including safety (nothing bad can happen in the future) and liveness (something good will eventually happen). For instance, the property ``if the sensor detects some movement, then the warning system must be activated'' is formalized in LTL as the formula $\square (detect\_motion \rightarrow \evt activate\_alert)$, where the modal operator $\square$ can be read as ``always'' and the modal operator $\evt$ as ``eventually in the future'' ($\rightarrow$ is ``implies'').

However, LTL cannot express that the warning system must be activated within some time frame---e.g., 10 seconds after motion has been detected. Precisely, MTL extends LTL by adding temporal constraints to the modal operators, thus allowing the specification of QTP \cite{Koymans1990,Ouaknine2007}.
In this way, the above requirement can be modeled by the following MTL formula: $\square (detect\_motion \rightarrow \evt_{[0,10]} activate\_alert)$, meaning that it is always the case that whenever $detect\_motion$ is true then $activate\_alert$ will eventually be true at most 10 seconds after $detect\_motion$ became true.

A slightly more involved example is the following: ``if the sensor detects some movement for at least 2 seconds, then the warning system must be activated no later than 10 seconds after that''. In this case the MTL formula is the following:
$\square(\square_{[0,2]} detect\_motion \rightarrow \evt_{(2,12]} activate\_alarm)$.
Here $\lozenge_{(2,12]}$ means that $activate\_alarm$ will hold in no more than 10 seconds after the sensor has detected movement for at least 2 seconds.

MTL is defined as follows.

\begin{definition}
Given a set of atomic predicates $P$, the MTL formulas are the expressions produced by the following grammar:
\[
\alpha ::= p \mid \lnot \alpha \mid \alpha \land \alpha \mid \alpha \lor \alpha \mid \alpha \cup_I \alpha
\]
where $p \in P$, $I \subseteq [0,\infty]$ is an integer interval\endnote{That is, $I$ has one of the following forms: $(a,b)$, $(a,b]$, $[a,b)$ and $[a,b]$ with $I \subseteq \nat \cup \{\infty\}$ such that $a$ cannot be $\infty$ and when $b$ is $\infty$ the interval must be open to the right.} and $U$ is the LTL modal operator called \emph{until}.
\end{definition}

A predicate such as $\phi\cup_I\alpha$ is true iff whenever $\phi$ becomes true it remains true until a time in $I$ in which $\phi$ does not hold anymore and $\alpha$ is true.

The following modal operators can be written as MTL predicates: $\evt_I \alpha \equiv  true \cup_I \alpha$ (called \emph{finally}); and $\Box_I \alpha \equiv \lnot \evt_I \lnot \alpha$ (called \emph{globally}). The until modal operator can be used without a subscript in which case it symbolizes $\cup_{[0, \infty)}$ (i.e., $\cup \equiv \cup_{[0, \infty)}$).

\subsection{Model Checking}
\label{Model Checking}

Model checking is a formal verification technique that exhaustively explores the state space of a model of a system in order to verify that a given property holds. Models are described in some formal language (e.g. automata, Petri nets, state machines), whereas properties are described in some logic (e.g. LTL, CTL). The algorithms or programs that explore the state space are called \emph{model checkers}. More formally, a model checker verifies that the language accepted by the model ($M$) is included in the language accepted by the property ($P$), noted $M \models P$. In other words, any behavior described by $M$ must be a behavior described by $P$. When $M \not\models P$ the model checker generates a counterexample. The counterexample is represented as a sequence of state transitions such that the property does not hold.
 
\begin{figure*}
	\begin{center}
		\includegraphics [scale=0.3]{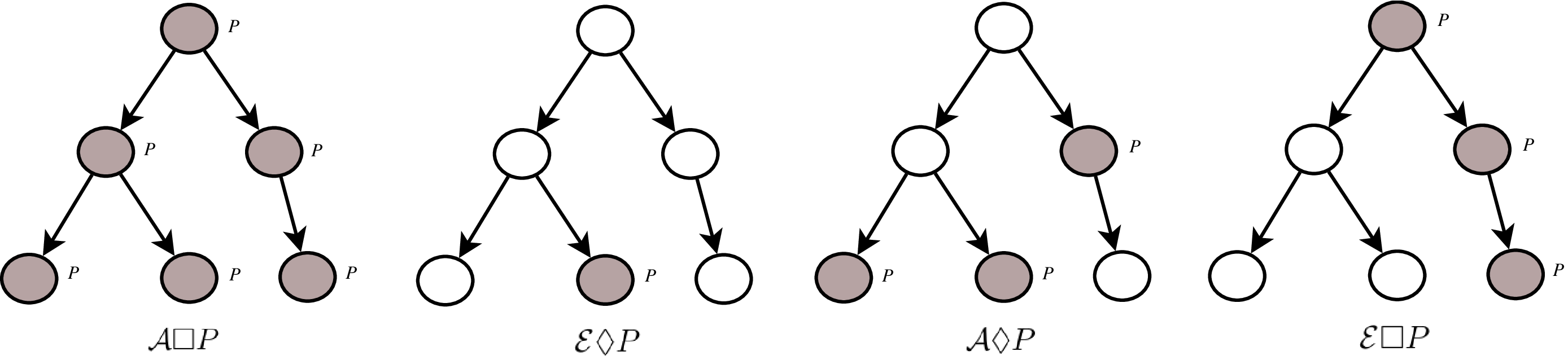}
	\end{center}
	\caption{TCTL formulas admitted by Uppaal}
	\label{fig:TCTL} 
\end{figure*}

Some model checkers admit models and properties of real-time systems. Uppaal \cite{Uppaal04} is one of the leading model checkers in academia and industry that can deal with certain classes of real-time problems. It uses TA for models and a TCTL subset for properties. 
The TCTL subset admitted by Uppaal does not allow timed versions of $\Box$ and $\evt$ as in MTL. Actually, only the following temporal formulas are supported by Uppaal\endnote{$\A$, $\Box$, $\E$ and $\evt$ are temporal modalities; $P$ is a predicate. The equivalent syntax in Uppaal is: $\A\equiv$ \texttt{A}, $\E\equiv$ \texttt{E}, $\Box\equiv$ \texttt{[]}, $\evt\equiv$ \texttt{<>}.}: 
$\A\Box P$ ($P$ always holds), $\E\evt P$ (there exists a state where $P$ holds), $\A\evt P$ ($P$ holds in at least one state of every execution path), and $\E\Box P$ ($P$ holds in all the states of at least one execution path), where $P$ is a predicate. These temporal formulas can be called \emph{path formulas}  because they quantify over paths or traces of states of the model. Figure \ref{fig:TCTL} illustrates the different path formulas supported by Uppaal. Gray (white) circles represent states where predicate $P$ (does not) holds, whereas edges link states in the same path. Then, for instance, as in the fourth graph $P$ holds in at least one path then $\E\Box P$ holds.

In summary, Uppaal works with a subset of TCTL which is different from MTL. MTL can describe QTR but it cannot distinguish execution paths, whereas the TCTL subset implemented by Uppaal cannot specify some recurrent QTR but it can specify what should happen in different execution paths---see Figure \ref{fig:TCTL}. In this work we show how TCTL combined with Uppaal's TA can be used to verify some important classes of QTP described with MTL. In order to achieve our goal we rely on a technique known as \emph{automata observer} \cite{MGT2009,BackesWGK16} that will be explained in Sections \ref{Implementación de los Patrones de propiedades temporales cuantitativas} and \ref{Verificar con Uppaal una propiedad}.

\paragraph{Timed Automata in Uppaal.}
Uppaal extends the TA introduced in Section \ref{Timed Automata} as follows \cite{Bengtsson2004}:
\begin{itemize}
\item \label{i:funbool} Boolean functions. Guards can include Boolean functions written in a language similar to the C programming language.
\item Channels and urgent channels. Channels are used to link and communicate two or more TA between each other. Urgent channels are used to fire an active transition immediately.
\item Committed nodes. When the system arrives to a committed node a transition must be fired immediately. This is useful to model sequences of atomic actions.
\item Shared variables. Allow for an asynchronous communication between TA. Shared variables can be integer variables or integer arrays with a bounded domain and an initial value. Predicates over these variables can be included in guards. A C-like syntax is used.
\end{itemize}

\section{Case Study}
\label{Caso de Estudio: Train-Gate}

We will use this case study as a running example in the rest of the paper. The case study is about a railway system that controls the access of $N$ trains to a bridge. We call it Railway Control System (RCS). Some of the requirements were introduced in order to have complex QTR. The bridge can be used by just one train. The system must enforce this restriction. Before the bridge there is the so-called crossing area. The access to the crossing area is regulated by a railway semaphore signal. This semaphore is controlled by the same system. When a train approaches the crossing area (\textsf{Appr}) the system can ask it to stop no later later than 10 seconds since its entrance, otherwise it is too late and the train must cross the bridge (\textsf{Cross}). If the train stops (\textsf{Stop}), later it restarts its journey (\textsf{Start}) in order to cross the bridge.

When a train enters the crossing area the controller asks it to stop if another train is crossing the bridge; otherwise it can cross the bridge. Trains in the crossing area wait in line until the controller orders them to move. If there are $N$ trains in the crossing area the red light of the semaphore is turned on. Once all the trains in the crossing area cross the bridge, the green light is turned on.

Besides, an alarm warns (\textsf{Warning}) the trains when the number  of trains in the crossing area is 3 or more. If after 2 seconds the number of trains does not decrease, the alarm will start to sound louder (\textsf{Danger}) in at most 5 seconds; otherwise the alarm is turned off (\textsf{Off}).
When the alarm sounds louder the system turns on the red light. The alarm will sound louder for at most 7 seconds after which it produces a different sound (\textsf{Howl}) which will last for at most 3 seconds unless the number of trains in the crossing area goes below 3; in either case the alarm is turned off. When the alarm is off the system turns on the green light.

If $N = 6$ the following QTP must hold:
\begin{enumerate}[label=(P\arabic*)]
\item\label{p1} If all the 6 trains are in the crossing area or on the bridge, all of them must leave the area before 122 seconds.
\item\label{p2} If there are at least 3 trains in the crossing area, the alarm must sound louder in at most 10 seconds.
\item\label{p3} If the alarm is sounding louder, the red light must be turned on for 9 seconds.
\end{enumerate}

\begin{figure*}
	\begin{center}
		\includegraphics [scale=0.35]{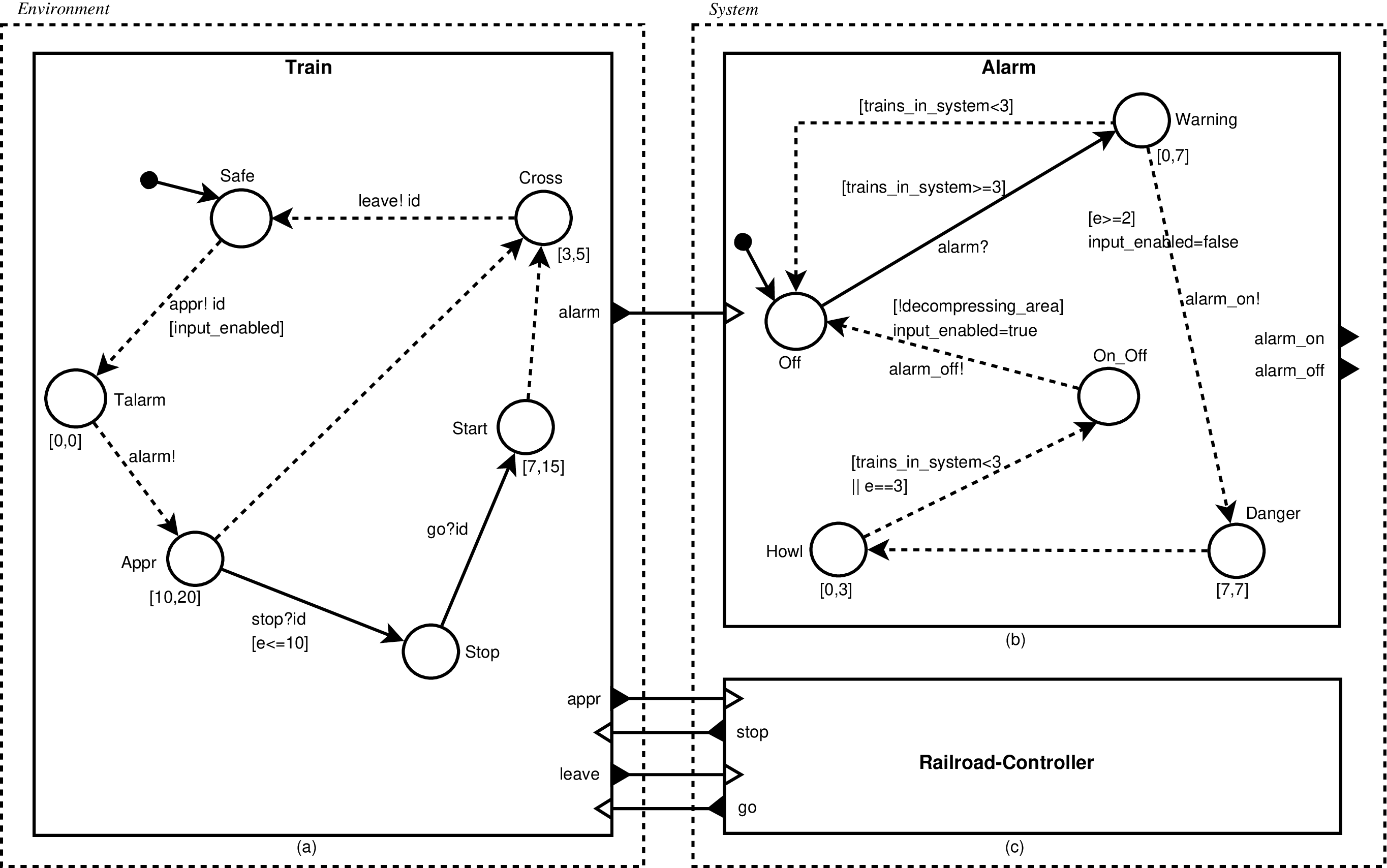}
	\end{center}
	\caption{RT-DEVS modeling the Railway Control System (RCS)}
	\label{fig:train-alarm-rtdevs} 
\end{figure*}

We model the RCS in RT-DEVS. RT-DEVS models can be represented in a simple graphic notation \cite{SK2005}, as can be seen in Figure \ref{fig:train-alarm-rtdevs}---Appendix \ref{RCS RT-DEVS} presents the RT-DEVS mathematical definition of the model. States are depicted as circles and their time intervals describe the value of function $t$---see Definition \ref{def:rt-devs} and recall we are using a simplified version of RT-DEVS. The initial state is identified by an arrow with a free origin. Internal transitions are represented by dotted line arrows. The label \texttt{p!v} over an internal transition represents the execution of the output function $\lambda$, where \texttt{p} is the output port and \texttt{v}  the value sent through it. External transitions are represented by solid line arrows. The label \texttt{p?v} represents the value coming in through port \texttt{p}. Both classes of transitions can be labeled with Boolean guards of the form \texttt{[expr]} that must be satisfied for the transition to occur. Coupled models are defined by drawing arrows linking the borders of the involved atomic models. The output ports are represented with solid triangles whereas the input ports are represented by empty triangles.

In Figure \ref{fig:train-alarm-rtdevs} the model named \textsf{Train} (figure (a)) describes the dynamics of the crossing area while the model called \textsf{Alarm} (figure (b)) describes the behavior of the alarm. The model \textsf{Railroad-Controller} (figure (c)) is not shown because it contains no time restrictions making it uninteresting for the present purposes.

The complete RT-DEVS specification and all the Uppaal code used in this paper can be found online in a GitHub repository \cite{github}.

\section{Verification of QTP Described with RT-DEVS and MTL}
\label{Metodologia de Verificacion de RT-DEVS}

In this work we describe a technique for the verification of some expressive, recurrent classes of QTP described with RT-DEVS and MTL, which can be decomposed in the following steps:
\begin{enumerate}
\item\label{i:rtdevsta} Translate the RT-DEVS models into TA (Section \ref{From RT-DEVS to TA}).
\item\label{i:prop} Express QTP using patterns of MTL formulas (Section \ref{Implementación de los Patrones de propiedades temporales cuantitativas}).
\item \label{i:mc} Use Uppaal to check if the TA of item \ref{i:rtdevsta} verify the QTP of item \ref{i:prop} (Section \ref{Verificar con Uppaal una propiedad}). 
\item \label{i:mut} If the RT-DEVS model fails to verify some QTP, apply specification mutation to the patterns of item \ref{i:prop} to find timing errors in the RT-DEVS model of item \ref{i:rtdevsta} (Section \ref{errorsModels}).

\item Use Uppaal and the mutants of item \ref{i:mut} to generate test cases to test the implementation of the RT-DEVS models (Section \ref{Generacion de casos de prueba}).
\end{enumerate}

We choose Uppaal for two reasons: (a) it is a mainstream model checker; and (b) the TA it supports combined with TCTL allow to express some important classes of QTP---those of item \ref{i:prop} above.

Next we develop each of the steps listed above using the RCS case study (Section \ref{Caso de Estudio: Train-Gate}) as a running example.

\subsection{From RT-DEVS to TA}
\label{From RT-DEVS to TA}

\begin{figure*}
     \centering
   \subfloat[TA Train]{\includegraphics [scale=0.29]{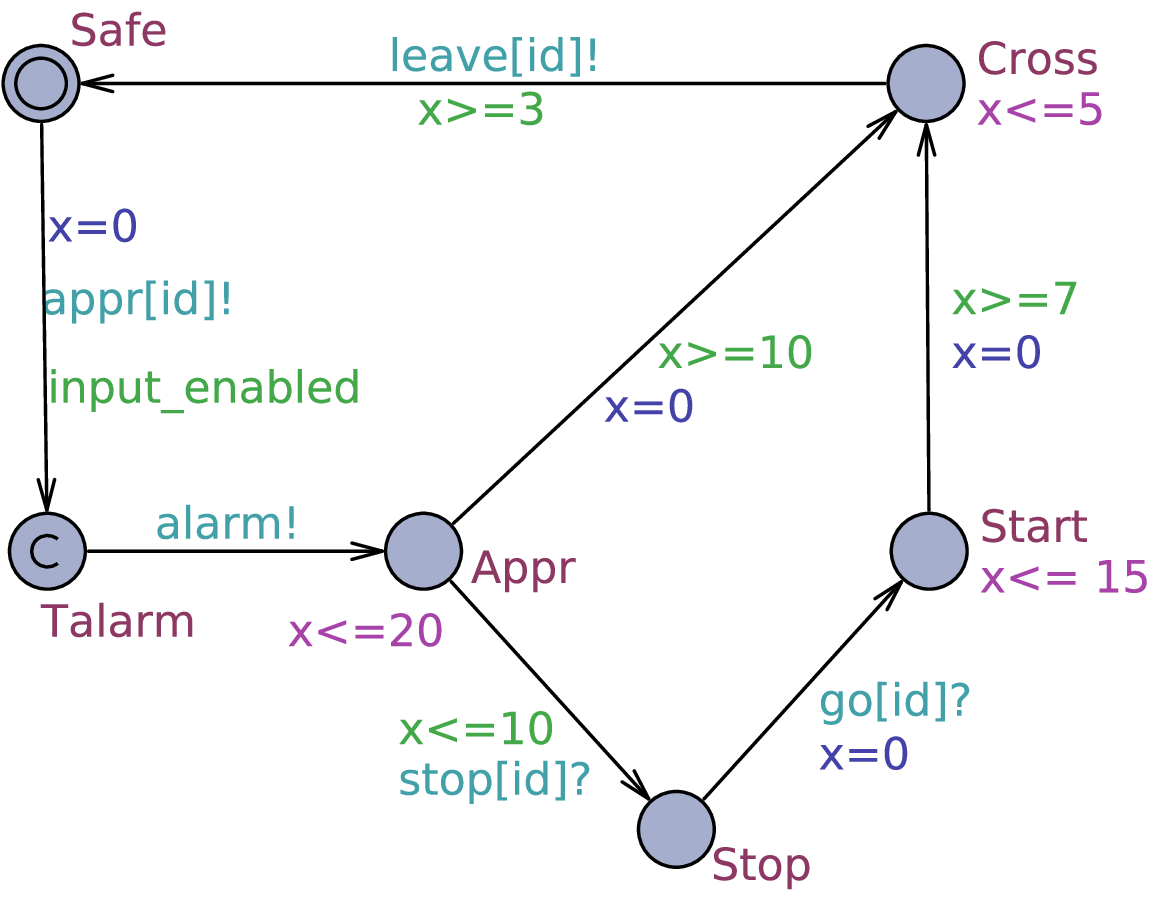}
      \label{fig:train}}
      \hfil
    \subfloat[TA Alarm]{\includegraphics [scale=0.44]{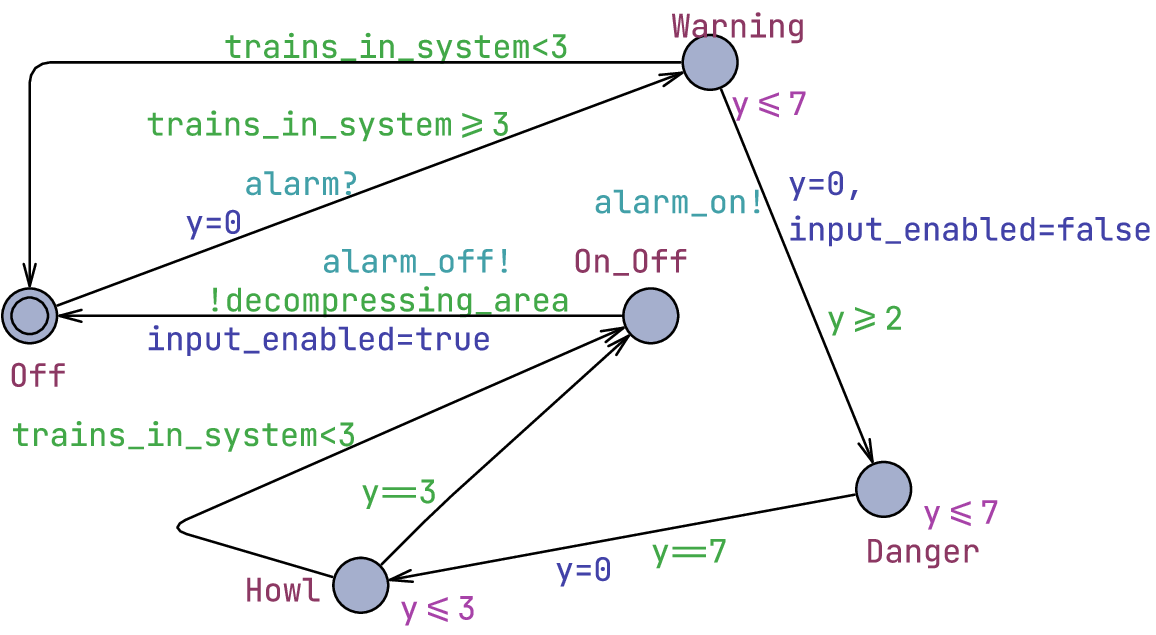}
      \label{fig:Alarm_extend_p_Cond_Sec}}
      \hfil

     \caption{TA resulting from the translation of the RT-DEVS models of Figure \ref{fig:train-alarm-rtdevs}}
     \label{fig:train-gate-alarm}
\end{figure*}

Several works propose different translations of variants of DEVS into TA \cite{SW2009,SaadawiW13,SWM12, FN2008}. We use exactly the same translation proposed by Furfano and Nigro \cite{SW2009}. This translation is described below using the \texttt{Train} and \texttt{Alarm} atomic RT-DEVS models depicted in Figure \ref{fig:train-alarm-rtdevs}, which results in the TA of Figure \ref{fig:train-gate-alarm}.

Currently, the translation is performed manually as proposed by Furfano and Nigro. Nevertheless, it could be implemented by a model transformation tool making part of a Model-Driven Development approach \cite{MCF03bis} Gonzáles et al. \cite{Gonzalez15,Gonzalez16} have developed model transformation tools taking DEVS models and producing PowerDevs code \cite{BK11}.

Each RT-DEVS atomic model is mapped to a TA. States in the RT-DEVS model are mapped to nodes with the same name in the TA. The initial node of a TA is depicted as a double line circle. Only one clock is defined in each TA to model the time flow in the RT-DEVS model.
RT-DEVS ports are modeled with channels included in transitions; whereas values sent and received through ports are modeled by means of shared variables.

Let $[a,b]$ be the time interval of some state in a RT-DEVS model, and let $x$ be the clock of the corresponding TA. This state is mapped to a TA node whose invariant is $x \leq b$ and where the transitions leaving it are guarded with the condition $a \leq x$. See, for instance, the \textsf{Cross} node and the transition towards the \textsf{Safe} node in Figure \ref{fig:train-alarm-rtdevs}(a).

An internal transition labeled with \textsf{p!v} is represented with a transition bound to the output channel \textsf{p!} which uses \textsf{v} as a shared variable. That is, if in the RT-DEVS model a value is sent from an output port to an input port, a shared variable in the resulting TA must be defined. As an example, see that variable \textsf{id} appearing in the transition labeled with \textsf{leave!id} in Figure \ref{fig:train-alarm-rtdevs}(a) corresponds to a variable of the same name in the TA of Figure \ref{fig:train-gate-alarm}(a). External transitions are translated much in the same way. When the value transmitted in a RT-DEVS is omitted there is no need to define a variable in the corresponding TA. This can be observed, by comparing the transition going from \textsf{Off} to \textsf{Warning} in Figure \ref{fig:train-alarm-rtdevs}(b) with the corresponding transition in  Figure \ref{fig:train-gate-alarm}(b). 

If in a RT-DEVS model we have $ti(s)=[0,0]$ for some state $s$, a committed state is used in the TA. For instance, \textsf{Talarm} in Figure \ref{fig:train-gate-alarm}(a) is a committed state as is marked with a \textsf{C} letter inside.

\subsection{Implementation of QTP patterns in Uppaal}
\label{Implementación de los Patrones de propiedades temporales cuantitativas}

QTP patterns capture an important class of recurrent temporal properties. Going from an informal statement of a temporal property to its formal statement in some temporal logic is usually a difficult task. Patterns of temporal properties help to reduce the gap between informal descriptions and temporal formulas \cite{DwyerAC99}. Each pattern informally describes a recurrent temporal property which is associated to a temporal formula where atomic predicates are given as placeholders. Engineers can look up an informal description and then they can substitute placeholders with specific atomic predicates. The pattern provides the temporal structure of the formula; users do not need to deeply understand the relation between, e.g., ``always'' and $\Box$, nor of more complex informal descriptions and combinations of the temporal modalities and logic connectives. Besides, since patterns capture many recurrent temporal properties, users only need to search the right formula among a handful of options. For example, the statement \emph{P must be followed by Q within k time-units} is the description associated to a pattern called \emph{Time-Bounded Response}, where $P$, $Q$ and $k$ are parameters that users must instantiate with predicates and numbers coming from their problems.

It is important to remark that we are not the first in proposing patterns for real-time properties but, as far as we know, we are the first in presenting a class of patterns of quantitative real-time properties that can be verified by a model checker. For example, Konrad and Cheng \cite{Konrad2005}, Gruhn and Laue \cite{GruhnL06}, and Abid et al. \cite{Abid2013} present patterns for real-time systems based on those introduced by Dwyer et al. \cite{DwyerAC99}, some of which produce formulas that cannot be verified by a model checker. After an analysis based on several works \cite{Ouaknine2006,Ouaknine2007,Ouaknine2008}, we have formalized in MTL a subset of the patterns whose formulas can be verified by a model checker. In the present paper some of these patterns are used to show the feasibility of the formal verification approach for RT-DEVS models.

In a previous work we have presented a collection of QTP patterns whose formulas are given in MTL \cite{SCCC2021}. Now we show how these patterns can be implemented in Uppaal in such a way that the tool can be used to check these properties on RT-DEVS models. The technique consists in defining a TA representing a QTP pattern and another TA representing the RT-DEVS model to be verified. The first TA is called \emph{observer} while the second is called \emph{model}. The observer and model TA communicate between each other while running in parallel. In this way, when the model TA transitions the observer TA controls if these transitions satisfy the property it represents or not. Consequently, verifying the property described by the pattern is reduced to some reachability problem about some \emph{bad} or \emph{error} state (if the property is not met) or some \emph{good} state (if the property is met) \cite{AcetoBL98,GruhnL06}. This technique is further explained in Section \ref{Verificar con Uppaal una propiedad}. The idea of using observers to capture temporal properties is widely used \cite{MGT2009,BackesWGK16, Abid2013}.

In the following sections we describe three QTP patterns supported by our technique. More patterns can be found in \ref{Patrones de propiedades temporales cuantitativas en Uppaal}.

\subsubsection{Time-Bounded Response.}
\label{Time-Bounded Response}

The time-bounded response property is perhaps the most recurrent property in real-time systems. This kind of properties are used to express a cause-effect relation such as $Q$ must eventually hold once $P$ has become true. In this particular case, the property not only forces an ordering on events but it also poses a time restriction on their occurrence. For instance, if a sensor detects some movement the alarm must sound before 3 seconds. The pattern is documented in Pattern \ref{patt:tbr}.

\begin{Pattern*}
\centerline{
\noindent
\fbox{%
\begin{tabularx}{.95\textwidth}{lX}
\textsc{Statement} & $P$ must be followed by $Q$ within $k$ time-units \\
\textsc{Description} & $Q$ holds as a response to $P$. The response must be given before $t(P)+k$, where $t(P)$ is the time on which $P$ became true.  \\
\textsc{MTL formula} & $\square (P \rightarrow \lozenge _{(0,k]} Q)$ \\
\textsc{Observer TA} & 
	\begin{center}
		\includegraphics [scale=0.4]{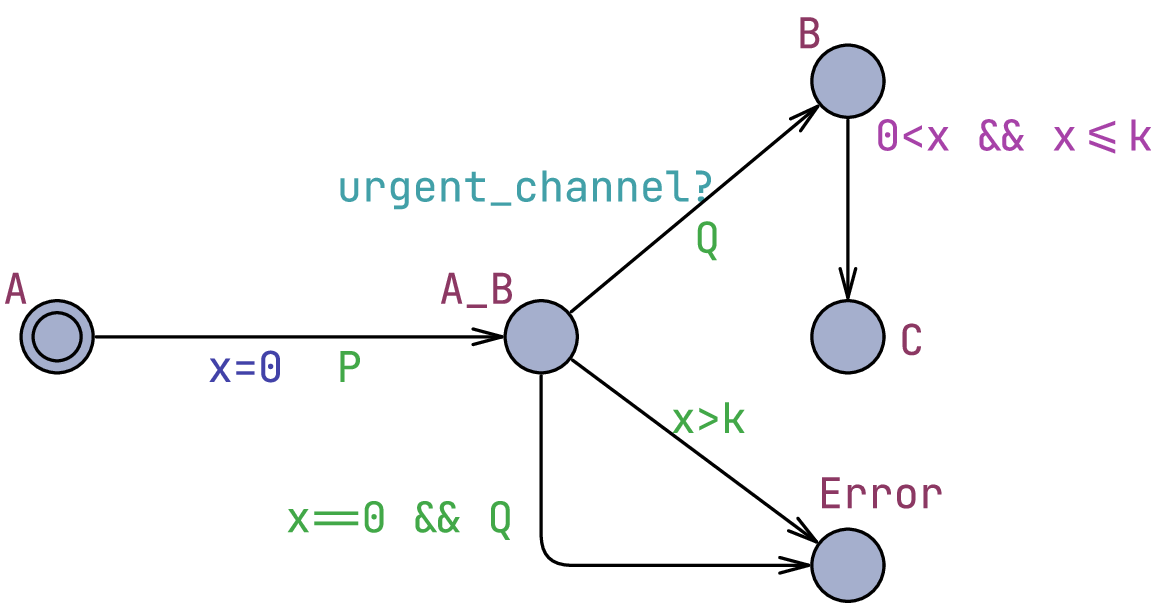}
	\end{center}
\end{tabularx}
}}
\caption{\label{patt:tbr}\emph{Time-Bounded Response}}
\end{Pattern*}

As can be seen, in the observer TA predicates $P$ and $Q$ are implemented as Boolean functions, although users can take advantage of other Uppaal features such as channels as we show in Section \ref{Application to the RCS}. Clock $x$ is used to control that $Q$ holds within the time interval $(0,k]$ counted from the moment that $P$ became true. See that $x$ is set to zero as soon as $P$ holds (transition from \textsf{A} to \textsf{A-B}). \textsf{Error} is a $bad$ state. 
\textsf{urgent\_channel} is a variable implementing an urgent channel thus forcing an instantaneous transition from \textsf{A-B} to \textsf{B} as soon as $Q$ becomes true.
Consequently, if a model TA transitions in such a way that the observer TA reaches \textsf{Error} it means that the property has been violated because after $k$ time units (t.u.) of $P$ becoming true, $Q$ remains false (straight transition from \textsf{A-B} to \textsf{Error}); or $P$ and $Q$ hold simultaneously (curved transition from \textsf{A-B} to \textsf{Error}).

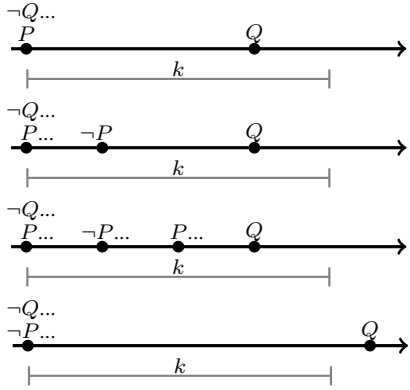
\begin{figure}
   \begin{center}
     \begin{tabular}{ l }

		\begin{tikzpicture} 
			{\draw [very thick,->] (-0.2,0) -- (5,0); \filldraw [black] (0,0) circle (2pt); \node (somenode) at (0.0,0.2) {\footnotesize $P$}; \node (noQ) at (0.06,0.47) {\footnotesize $\neg Q$...}; \filldraw [black] (3,0) circle (2pt); \node  at (3,0.2) {\footnotesize $Q$}; \node  (intervalo) at (2,-0.26) {\footnotesize  $k$}; \draw [thick, |-|,gray] (0,-0.4) -- (4,-0.4); }
		\end{tikzpicture}  
\\  
		\begin{tikzpicture} 
			{\draw [very thick,->] (-0.2,0) -- (5,0); \filldraw [black] (0,0) circle (2pt); \node (somenode) at (0.14,0.2) {\footnotesize $P$...}; \node (noQ) at (0.06,0.47) {\footnotesize $\neg Q$...}; \filldraw [black] (3,0) circle (2pt); \node  at (3,0.2) {\footnotesize $Q$}; \filldraw [black] (1,0) circle (2pt); \node  at (0.92,0.2) {\footnotesize $\neg P$}; \node  (intervalo) at (2,-0.26) {\footnotesize $k$}; \draw [thick, |-|,gray] (0,-0.4) -- (4,-0.4); }
		\end{tikzpicture}
\\
		\begin{tikzpicture} 
			{\draw [very thick,->] (-0.2,0) -- (5,0); \filldraw [black] (0,0) circle (2pt); \node (somenode) at (0.14,0.2) {\footnotesize $P$...}; \node (noQ) at (0.06,0.47) {\footnotesize $\neg Q$...}; \filldraw [black] (3,0) circle (2pt); \node  at (3,0.2) {\footnotesize $Q$}; \filldraw [black] (1,0) circle (2pt); \node  at (1.04,0.2) {\footnotesize $\neg P$...};  \filldraw [black] (2,0) circle (2pt); \node  at (2.11,0.2) {\footnotesize $P$...}; \node  (intervalo) at (2,-0.26) {\footnotesize $k$}; \draw [thick, |-|,gray] (0,-0.4) -- (4,-0.4); }			
		\end{tikzpicture}
 \\  
		\begin{tikzpicture} 
			{\draw [very thick,->] (-0.2,0) -- (5,0); \filldraw [black] (0,0) circle (2pt); \node (somenode) at (0.04,0.2) {\footnotesize $\neg P$...}; \node (noQ) at (0.04,0.47) {\footnotesize $\neg Q$...};\filldraw [black] (4.5,0) circle (2pt); \node  at (4.5,0.2) {\footnotesize $Q$};  \node  (intervalo) at (2,-0.26) {\footnotesize $k$}; \draw [thick, |-|,gray] (0,-0.4) -- (4,-0.4); }			
		\end{tikzpicture}
 \end{tabular} 	

   \end{center}
   \caption{Traces of the \emph{Time-Bounded Response} pattern not reaching \textsf{Error}}
   \label{tikz:Trazas_TB_Response} 
\end{figure}

Figure \ref{tikz:Trazas_TB_Response} shows some traces that make the observer TA not to reach \textsf{Error}; that is, traces verifying the property. These traces can help the user to find out if their problem is an instance of this pattern. The notation is as follows:
\begin{tikzpicture} {\draw [very thick] (-0.2,0) -- (0.6,0); \filldraw [black] (0,0) circle (2pt); \node [scale=0.7] (somenode) at (0.0,0.2) {\footnotesize $P$};}
\end{tikzpicture}, $P$ holds in that point in time;
\begin{tikzpicture} {\draw [very thick] (-0.2,0) -- (0.6,0); \filldraw [black] (0,0) circle (2pt); \node [scale=0.7] (somenode) at (0.1,0.2) {\footnotesize $P$...};}
\end{tikzpicture}, $P$ holds from that point and on; and
\begin{tikzpicture} {\node  [scale=0.7] (intervalo) at (0.45,0.15) {\footnotesize  $k$}; \draw [thick, |-|,gray] (0,0) -- (0.9,0); }\end{tikzpicture}, represents a time interval of length $k$.

\paragraph{Application to the RCS.}
Property \ref{p1} of the RCS can be put in terms of the \emph{Time-Bounded Response} pattern by instantiating $P$, $Q$ and $k$ as follows:
\begin{itemize}[leftmargin=*]
\item \textsf{trains\_in\_system == N} must be followed by \textsf{trains\_in\_system == 0} within \textsf{122} t.u.
\item MTL: $\Box (\textsf{trains\_in\_system == N}$ \\
\mbox{}\hspace{1.5cm}$\rightarrow \evt _{(0,122]}$ $\textsf{trains\_in\_system == 0})$
\end{itemize}
where \textsf{trains\_in\_system} is the variable used in Figure \ref{fig:train-gate-alarm}(b). In addition, the observer TA is instantiated by substituting \textsf{k} with \textsf{122}, and providing the following Uppaal implementations for $P$ and $Q$:
\begin{verbatim}
bool P() {return train_in_system == N;}
bool Q() {return train_in_system == 0;}
\end{verbatim}

\subsubsection{Time-Restricted Precedence.}

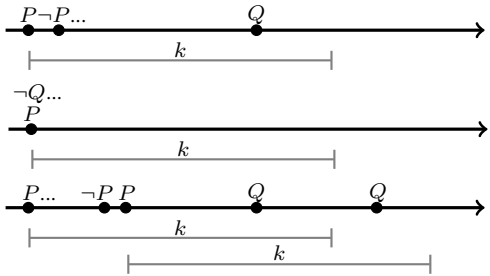
\begin{figure}
   \begin{center}
     \begin{tabular}{ l }                   
		\begin{tikzpicture} 
			{\draw [very thick,->] (-0.3,0) -- (6,0); \filldraw [black] (0,0) circle (2pt); \filldraw [black] (0.4,0) circle (2pt); \node (somenode) at (0.32,0.2) {\footnotesize $P  \neg P$...}; \filldraw [black] (3,0) circle (2pt); \node  at (3,0.2) {\footnotesize $Q$}; \node  (intervalo) at (2,-0.26) {\footnotesize $k$}; \draw [thick, |-|,gray] (0,-0.4) -- (4,-0.4);}
		\end{tikzpicture}  
\\
		\begin{tikzpicture} 
			{\draw [very thick,->] (-0.3,0) -- (6,0); \filldraw [black] (0,0) circle (2pt); \node (somenode) at (0.01,0.2) {\footnotesize $P$}; \node (noQ) at (0.08,0.47) {\footnotesize $\neg Q$...}; \node  (intervalo) at (2,-0.26) {\footnotesize $k$}; \draw [thick, |-|,gray] (0,-0.4) -- (4,-0.4); }			
		\end{tikzpicture}
\\
		\begin{tikzpicture}
 		{\draw [very thick,->] (-0.3,0) -- (6,0); \filldraw [black] (0,0) circle (2pt); \node (somenode) at (0.14,0.2) {\footnotesize $P$...}; 
 		 \filldraw [black] (1,0) circle (2pt); \filldraw [black] (1.28,0) circle (2pt); \node  at (1.05,0.2) {\footnotesize $\neg P \: P$};
 		 \filldraw [black] (3,0) circle (2pt); \node  at (3,0.2) {\footnotesize $Q$};
 		  \node  at (4.6,0.2) {\footnotesize $Q$}; \filldraw [black] (4.58,0) circle (2pt); \node  (intervalo) at (2,-0.26) {\footnotesize $k$}; \draw [thick, |-|,gray] (0,-0.4) -- (4,-0.4);  \node  (intervalo2) at (3+0.3,-0.6) {\footnotesize $k$}; \draw [thick, |-|,gray] (1+0.3,-0.75) -- (5+0.3,-0.75); }

		\end{tikzpicture}	
  \end{tabular} 	

   \end{center}
   \caption{Traces of the \emph{Time-Restricted Precedence} pattern not reaching \textsf{Error}}
   \label{tikz:Trazas_TR_Precedence} 
\end{figure}

Precedence properties refer to situations where $P$ allows $Q$ to hold in some time interval. In other words, $Q$ cannot hold if $P$ did not hold $k$ t.u. before. 
For instance, a secure door can only be opened if an access card was inserted 20 seconds before as the earliest.
Note that $Q$ is not \emph{forced} to hold after $P$ became true; $Q$ \emph{may} hold after $P$ became true.
Hence, if the secure door is not opened the precedence property still holds.
The pattern is documented in Pattern \ref{patt:trp} and some traces not reaching the \textsf{Error} state are shown in Figure \ref{tikz:Trazas_TR_Precedence}.

\begin{Pattern*}
\centerline{
\noindent
\fbox{%
\begin{tabularx}{.95\textwidth}{lX}
\textsc{Statement} & $P$ enables $Q$ for $k$ time-units \\
\textsc{Description} & $Q$ can hold only if $P$ holds and the elapsed time is not beyond $t(P)+k$. If $Q$ holds after $k$ time units since $P$ started to be true, the property does not hold.  \\
\textsc{MTL formula} & $\evt_{(0,k]} Q \rightarrow (\neg Q \cup  _{[0,k)} P)$ \\
\textsc{Observer TA} &
	\begin{center}
		\includegraphics [scale=0.47]{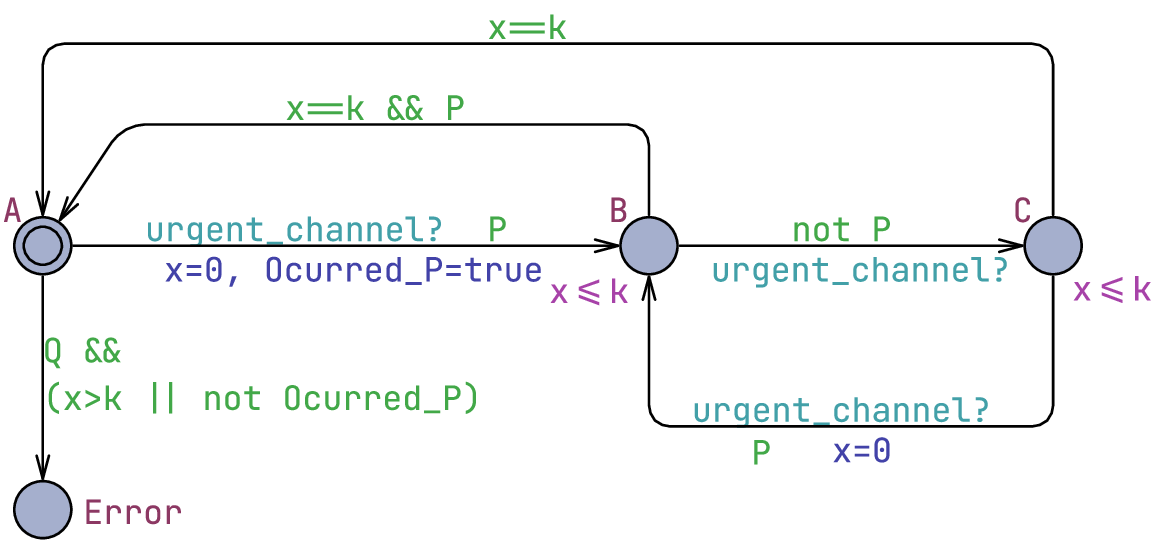}
	\end{center}
\end{tabularx}
}}
\caption{\label{patt:trp}\emph{Time-Restricted Precedence}}
\end{Pattern*}

\paragraph{Application to the RCS.}\label{Application to the RCS}
Property \ref{p2} of the RCS can be put in terms of the \emph{Time-Restricted Precedence} pattern by instantiating $P$, $Q$ and $k$ as follows:
\begin{itemize}[leftmargin=*]
\item \textsf{trains\_in\_system $\geq$ 3} enables \textsf{alarm\_on?} for \textsf{10} time-units
\item MTL: $\evt_{(0,10]} \textsf{alarm\_on?}$ \\
\mbox{}\hspace{1cm}$\rightarrow (\neg (\textsf{alarm\_on?}) \cup_{[0,10)} \textsf{trains\_in\_system $\geq$ 3})$
\end{itemize}

In this case, instead of implementing $Q$ with a Boolean function, it is implemented with the input channel \textsf{alarm\_on?}. $Q$ holds when the channel receives a signal from the corresponding output channel \textsf{alarm\_on!} present in the \textsf{Alarm} TA.

\subsubsection{Conditional Security.}
\label{Conditional Security}

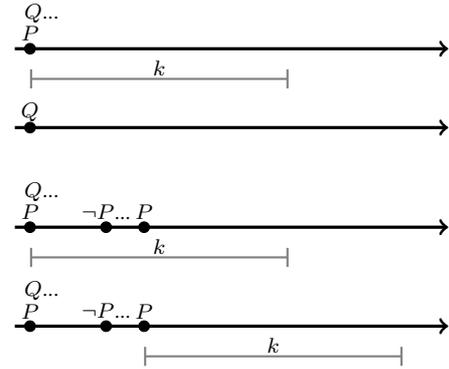
\begin{figure}
   \begin{center}
     \begin{tabular}{ l }                   
		\begin{tikzpicture} 
			{\draw [very thick,->] (-0.2,0) -- (5.5,0); \filldraw [black] (0,0) circle (2pt); \node (primerP) at (0,0.2) {\footnotesize $P$}; \node (primerQ) at (0.15,0.47) {\footnotesize $Q$...}; \node  (intervalo) at (1.7,-0.26) {\footnotesize $k$}; \draw [thick, |-|,gray] (0,-0.4) -- (3.4,-0.4); 
			}
		\end{tikzpicture}  
\\ 
		\begin{tikzpicture} 
			{\draw [very thick,->] (-0.2,0) -- (5.5,0);  
			\node (primerP) at (0,0.2) {\footnotesize $Q$}; \filldraw [black] (0,0) circle (2pt);  
			\draw [white] (0,-0.28) rectangle  (4,-0.48); }
		\end{tikzpicture}  
\\
		\begin{tikzpicture} 
			{\draw [very thick,->] (-0.2,0) -- (5.5,0);  \node (primerP) at (0,0.2) {\footnotesize $P$}; \filldraw [black] (0,0) circle (2pt);  \node (primerQ) at (0.15,0.47) {\footnotesize $Q$...}; \filldraw [black] (1,0) circle (2pt); \node (noP) at (1,0.2) {\footnotesize $\neg P$...}; \filldraw [black] (1.5,0) circle (2pt); \node (segP) at (1.5,0.2) {\footnotesize $P$}; \node  (intervalo) at (1.7,-0.26) {\footnotesize $k$}; \draw [thick, |-|,gray] (0,-0.4) -- (3.4,-0.4);  }
		\end{tikzpicture}  
\\
		\begin{tikzpicture} 
			{\draw [very thick,->] (-0.2,0) -- (5.5,0);  
			\node (primerP) at (0,0.2) {\footnotesize $P$}; \filldraw [black] (0,0) circle (2pt);  \node (primerQ) at (0.15,0.47) {\footnotesize $Q$...}; \filldraw [black] (1,0) circle (2pt); \node (noP) at (1,0.2) {\footnotesize $\neg P$...}; \filldraw [black] (1.5,0) circle (2pt); \node (segP) at (1.5,0.2) {\footnotesize $P$};
			 \node  (intervalo2) at (3.2,-0.26) {\footnotesize $k$}; \draw [thick, |-|,gray] (1.5,-0.4) -- (4.9,-0.4); }
		\end{tikzpicture}  

  \end{tabular} 	
  \end{center}
   \caption{Traces of the \emph{Conditional Security} pattern not reaching \textsf{Error}}
   \label{tikz:Trazas_Conditional_Security} 
\end{figure}

This property deals with situations where a predicate \emph{must} hold for a time interval counted from the moment another predicate became true. In particular if the latter remains true while the former is true, the time interval is extended accordingly. For example, if a sensor detects some movement, an indicator light must be turned on and kept on for 10 seconds. If, while the light is on, the sensor detects new movement, the light must remain on for 10 seconds after the second activation of the sensor. The pattern is documented in Pattern \ref{patt:cs} and some traces not reaching \textsf{Error} are shown in Figure \ref{tikz:Trazas_Conditional_Security}.

\begin{Pattern*}
\centerline{
\noindent
\fbox{%
\begin{tabularx}{.95\textwidth}{lX}
\textsc{Statement} & If $P$ then $Q$ holds for $k$ time-units \\
\textsc{Description} & This property states that if $P$ holds $Q$ must hold during $k$ time units since $P$ became true.  \\
\textsc{MTL formula} & $\Box (P \rightarrow \Box_{[0,k]} Q)$ \\
\textsc{Observer TA} &
	\begin{center}
		\includegraphics [scale=0.42]{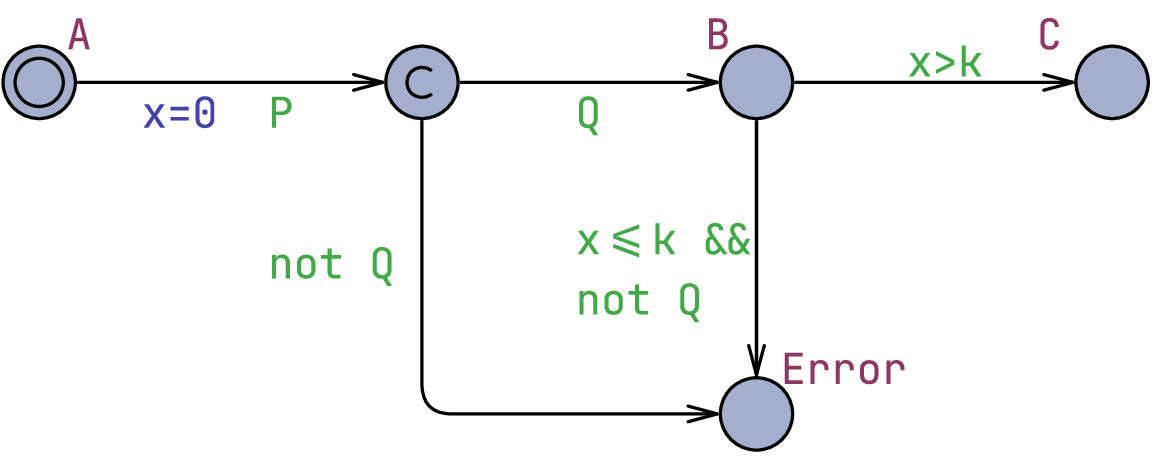}
	\end{center}
\end{tabularx}
}}
\caption{\label{patt:cs}\emph{Conditional Security} }
\end{Pattern*}

\paragraph{Application to the RCS.}
Property \ref{p3} of the RCS can be put in terms of the \emph{Conditional Security} pattern by instantiating $P$, $Q$ and $k$ as follows:
\begin{itemize}[leftmargin=*]
\item If \textsf{alarm\_on?} then \textsf{input\_enabled == false} holds for \textsf{9} time-units
\item MTL: $\Box (\textsf{alarm\_on?} \rightarrow \Box_{[0,9]} \textsf{input\_enabled == false})$
\end{itemize}
where \textsf{input\_enabled} is a variable used in Figure \ref{fig:train-gate-alarm}(a) and (b).

\subsection{Verification of QTP with Uppaal}
\label{Verificar con Uppaal una propiedad}

\begin{figure*}[!h] 
	\begin{center}
		\includegraphics [scale=0.5]{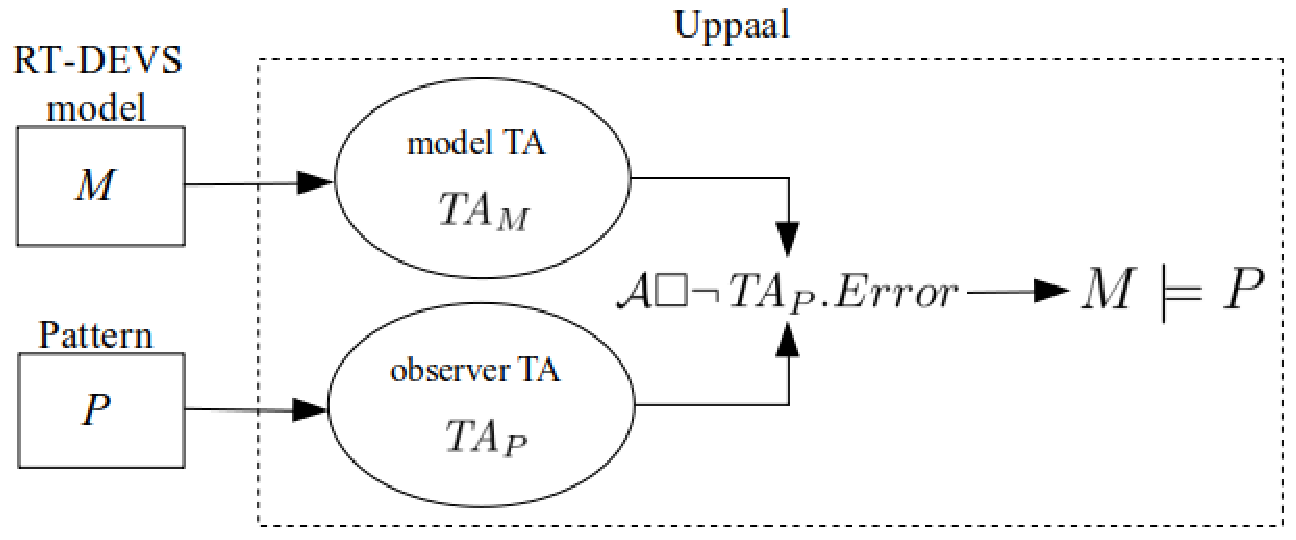}
	\end{center}
	\caption{Using Uppaal to verify QTP}
	\label{fig:verificar-Uppaal}
\end{figure*}

The technique to check if a RT-DEVS model ($M$) verifies a QTP ($P$) instantiated from one of the patterns described above, is depicted in Figure \ref{fig:verificar-Uppaal}. $M$ is translated into a TA, called $\mathit{TA}_M$, as shown in Section \ref{From RT-DEVS to TA}. The TA of the pattern from which $P$ comes from is instantiated accordingly, called $\mathit{TA}_P$. $\mathit{TA}_M$ corresponds to the model TA while $\mathit{TA}_P$ corresponds to the observer TA---see Section \ref{Implementación de los Patrones de propiedades temporales cuantitativas}. As we already said, these two TA are executed concurrently in such a way that the latter transitions as a response to the transitions performed by the former. On the other hand, the TCTL formula $\A\Box\lnot \mathit{TA}_P.\mathsf{Error}$ states that $\mathit{TA}_P$ never reaches the \textsf{Error} state in any of the execution paths. Hence, if that TCTL query holds $M$ verifies $P$, otherwise it does not. The TCTL query $\E\evt\mathit{TA}_P.\mathsf{Error}$ is an alternative way of concluding the same but in a dual way: if this query holds, the model does not verify the property, otherwise it does. 

The properties considered in the RCS case study are verified with the above technique taking $N=6$. In doing so, six instances of the TA named \textsf{Train}, one of the TA named \textsf{Alarm} and \textsf{Railroad-Controller} are executed concurrently.
Table \ref{table:verificarpatron} shows how each pattern is instantiated and the result of executing the TCTL query $\A\Box\lnot \mathit{TA}_P.\mathsf{Error}$ on Uppaal.
As can be seen, there are two properties that cannot be verified. In the next section we will explore a technique that helps in finding the causes of these errors.

\begin{table*}[ht]
\small\sf\centering
\caption{Verification of the RCS properties with $N=6$}
\label{table:verificarpatron}
\begin{tabular}{ c c c c c }
\toprule
$N=6$ & $k$ & $P$ & $Q$ & $M \models P$\\
\midrule
\ref{p1}&122& \textsf{trains\_in\_system == N} & \textsf{trains\_in\_system == 0} & no \\
\ref{p2}&10&\textsf{trains\_in\_system $\geq$ 3} &  \textsf{alarm\_on?} & yes \\
\ref{p3}&9 & \textsf{alarm\_on?} & \textsf{input\_enabled==false} & no \\
\bottomrule
\end{tabular} 
\end{table*}

\section{Finding timing errors with mutants of patterns of QTP}
\label{Encontrado errores con mutantes de patrones temporales}

Timing errors in real-time systems are hard to detect. Model checkers help in this regard because they return a counterexample whenever the model does not verify a certain property. Then, we know there is something wrong with the model and we have a witness of that. However, these counterexamples may be hard to understand and analyze. For example, the counterexample returned by Uppaal when it fails to prove property \ref{p1} has 35 transitions. Analyzing counterexamples of industrial-grade models can be overwhelming \cite{Counterexamples2009,Debbi2018Counterexamples,Kaleeswaran2022}.
For this reason in this section we propose a complementary technique to find quantitative timing errors in RT-DEVS models. If the model has timing errors, a possible interpretation is to think that it verifies a different temporal property. Consider a RT-DEVS model $M$ that should verify some QTP $P$, but it does not. Our conjecture is that it verifies a slightly different property $P'$, called a \emph{mutant} of $P$. If $M$ verifies $P'$ (i.e. $M \models P'$) the error has been detected. In this case we do not need a counterexample because we have found out what is the error in the model. This idea is inspired in mutation testing \cite{Buchler11,TanSL04,Trakhtenbrot17} and specification mutation \cite{DBLP:conf/issta/StocksC93}.

In a previous work we have shown how mutants of QTP can be generated \cite{SCCC2021}. These mutants are obtained by working at the pattern level---cf. Section \ref{Implementación de los Patrones de propiedades temporales cuantitativas}. Hence, if $P$ is an instance of a pattern,  users can use the mutants defined for that pattern. The mutants of our proposal are \emph{semantic mutants} meaning that they embody interpretation errors rather than simple syntactic errors.
In this paper we extend our previous results by showing how these mutants can be implemented in Uppaal by mutating the observer TA of patterns or by changing the TCTL query---see Section \ref{errorsModels}.

Once the cause of an error is found (i.e. why \ref{p1} fails), the RT-DEVS model or the requirements can be modified in such a way that the new version verifies the intended properties.
Besides, when the model verifies all the intended properties (i.e. the model is correct), mutants and model checking are still useful to generate test cases to test the implementation---see Section \ref{errorsImp}.

\subsection{\label{errorsModels}Finding timing errors in RT-DEVS models}
Next, we illustrate our technique with examples taken from the RCS case study. In Table \ref{table:verificarpatron} we can see that the RT-DEVS model does not verify properties \ref{p1} and \ref{p3} when $N = 6$. Then, we analyze possible errors in the model by considering mutants of \ref{p1}; a similar analysis can be performed over \ref{p3}.

\begin{figure}
   \begin{center}
		\begin{tikzpicture} 
			{\draw [very thick,->] (-0.2,0) -- (6.5,0); \filldraw [black] (0,0) circle (2pt); \node (primerP) at (0,0.2) {\footnotesize $P$};  \node (noQ) at (0.06,0.47) {\footnotesize $\neg Q$...};  \filldraw [black] (4.2,0) circle (2pt); \node (Q) at (4.2,0.2) {\footnotesize $Q$}; \node  (intervalo) at (1.7,-0.26) {\footnotesize  $k$}; \draw [thick, |-|,gray] (0,-0.4) -- (3.4,-0.4);  \node  (intervalo2) at (2.7,-0.6) {\footnotesize  $k'$}; \draw [thick, |-|,gray] (0,-0.75) -- (4.9,-0.75); }
		\end{tikzpicture}  

   \end{center}
   \caption{Trace satisfying \emph{Mutant 1}}
   \label{tikz:Trazas_TA_tb_1} 
\end{figure}
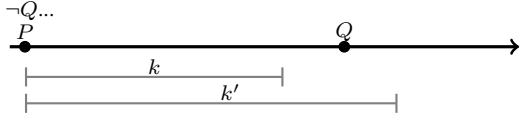

\paragraph{Mutant 1.} 
Would \ref{p1} hold with a larger $k$? This mutant serves to check if response $Q$ eventually holds but in a later time, as depicted in Figure \ref{tikz:Trazas_TA_tb_1}. Formally, the mutant corresponds to the MTL formula: $\Box (P \rightarrow \evt_{(0,k']} Q)$, with $k'>k$. That is, the time interval $(0,k]$ is extended to a larger one.

For instance, we could check if all the trains can leave the crossing area in 126 seconds, instead of the original 122. It can be done by instantiating the TA of Pattern \ref{patt:tbr} with $k=126$. Then, run on Uppaal $\A\Box\lnot \mathit{TA}'_{tb1}.\mathsf{Error}$ where $\mathit{TA}'_{tb1}$ is the mutant TA. In this case Uppaal answers that $M$ verifies the mutant property. Hence, the user can figure out what the problem with $M$ is with this new information. A possible error could be the time trains need to transit throughout the crossing area and the bridge. For instance, if crossing the bridge takes between 3 and 4 seconds (i.e. setting $ti(\mathsf{Train.Cross}) = [3,4]$ in Figure \ref{fig:train-alarm-rtdevs}, instead of $[3,5]$), \ref{p1} will hold.

\paragraph{Mutant 2.}
Is there a state trace where $Q$ holds in at most $k$ t.u.? Formally, this mutant is described by the following MTL formula: $\evt (P \rightarrow \evt_{(0,k]} Q)$. That is, the leading $\square$ is substituted by $\lozenge$ in the pattern formula. This modification captures a weaker interpretation of the property which holds if there is at least one state trace in which the answer arrives on time. Hence, in a way, we are assuming that the engineer misinterpreted a requirement or property.

To answer this question we verify if a \emph{good} state is reachable. That is, we first instantiate the TA of Pattern \ref{patt:tbr} as in Section \ref{Time-Bounded Response} (i.e. with  $k=122$), but then we run the following query: $\E\evt \mathit{TA}_{tb2}.\mathsf{B}$, where $\mathit{TA}_{tb2}$ is the instance of the pattern. Note that in this case we do not introduce a mutation in the TA but in the TCTL query. In other words, the mutation is implemented by running an adequate reachability query. 

If the answer to this query is negative, it means that $Q$ holds too late. In the RCS example there is at least one way where all the 6 trains free the crossing area in at most 122 seconds. However, if $k$ is set to 40 seconds, it is impossible for all the 6 trains to leave on time.

\paragraph{Mutant 3.}
Can $P$ and $Q$ hold at the same time? This situation does not satisfy the \emph{Time-Bounded Response} pattern. Formally, this mutant is expressed in MTL as follows: $\evt (P \rightarrow \evt_{[0,0]} Q)$. That is $Q$ must hold as soon as $P$ becomes true.

\begin{figure}
    \centering
    \begin{subfigure}[t]{0.3\textwidth}
        \caption{ } 
        \includegraphics[scale=0.28]{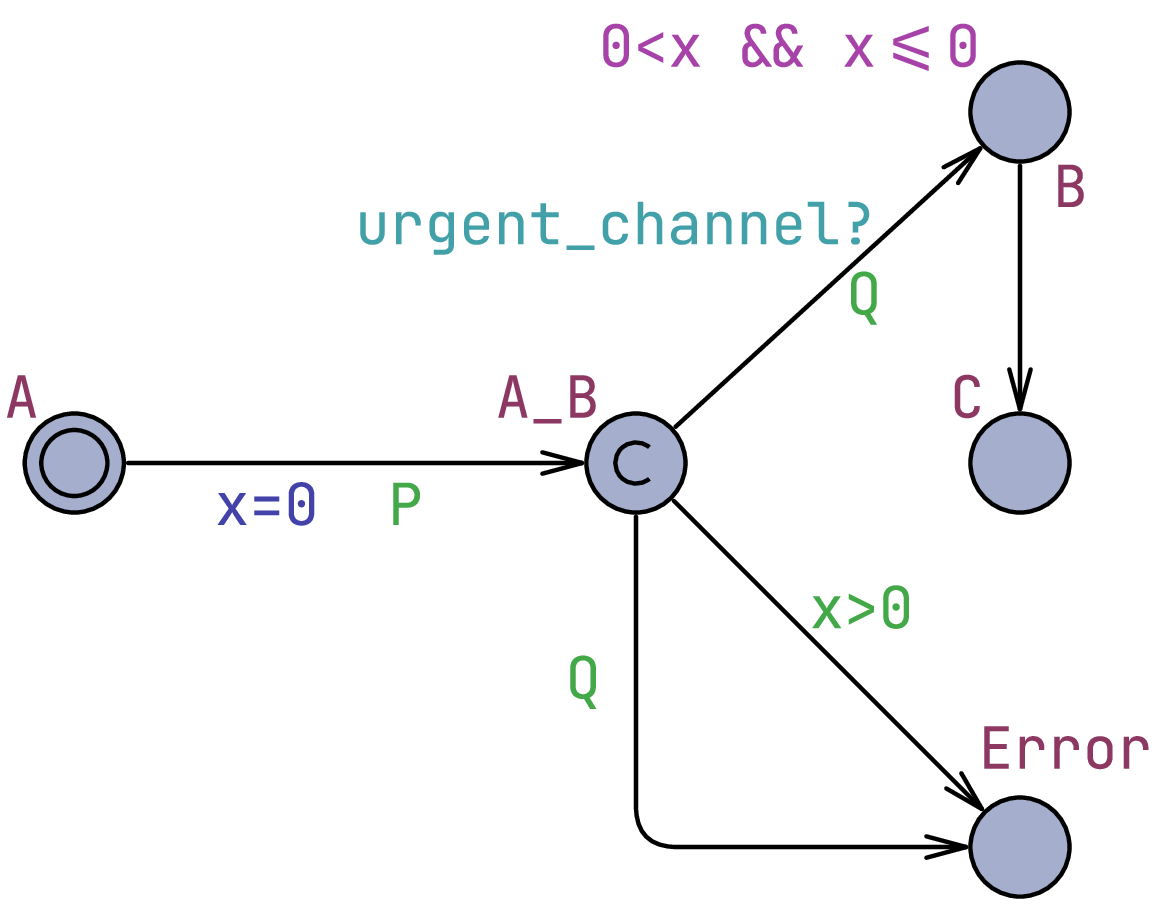}
        \label{fig:mut-3-TBR}
    \end{subfigure}
    \hfill
    \begin{subfigure}[t]{0.3\textwidth}
        \caption{}
        \includegraphics[scale=0.27]{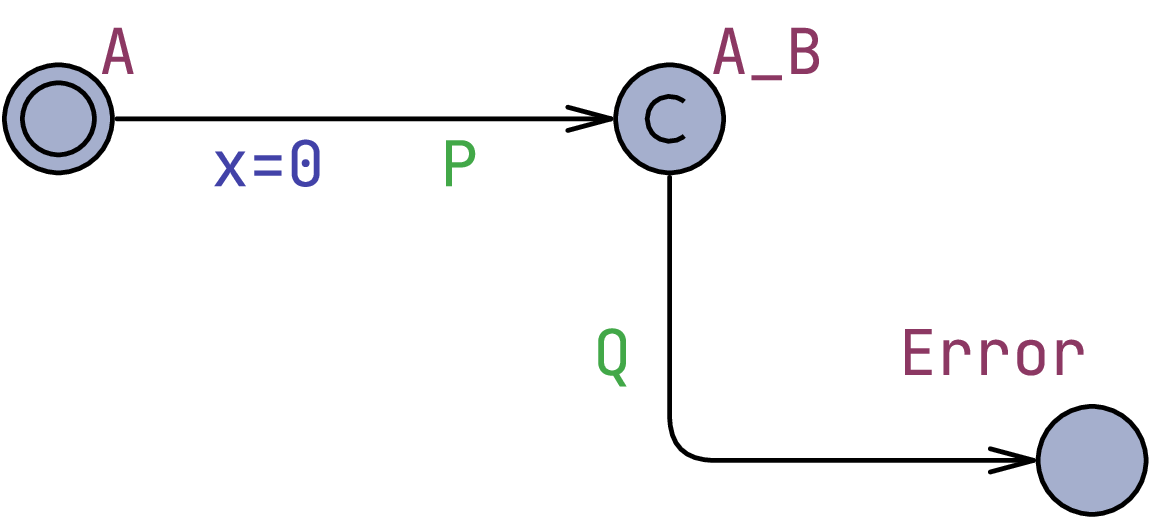}
        \label{fig:mut-3-TBR-simp} 
    \end{subfigure}
    \hfill
    \begin{subfigure}[t]{0.3\textwidth}
        \caption{}
        \includegraphics[scale=0.27]{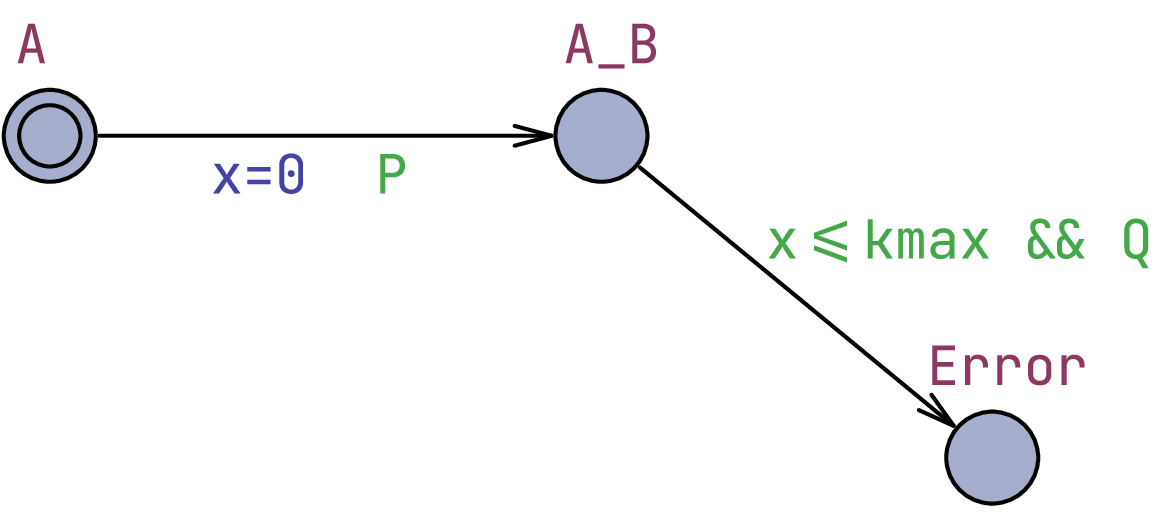}
        \label{fig:mut-4-TBR} 
    \end{subfigure}
    
    \caption{\label{mut34} (a) TA of \emph{Mutant 3}; (b) Simplified TA of \emph{Mutant 3}; and  (c) TA of \emph{Mutant 4}}
\end{figure}

The TA of Pattern \ref{patt:tbr} thus mutates into $\mathit{TA}'_{tb3}$, as depicted in Figure \ref{mut34}(\subref{fig:mut-3-TBR}). As can be seen,  the \textsf{A-B} state becomes a committed state, and $k$ is set to zero.
As we have already explained, a committed state forces Uppaal to leave it immediately. Hence, in this case, if $Q$ does not hold at the moment \textsf{A-B} is reached the \textsf{Error} state is never reached.
However, this TA can be simplified to the TA shown in Figure \ref{mut34}(\subref{fig:mut-3-TBR-simp}).
Since the invariant of node \textsf{B} is impossible to be satisfied the fragment of the TA comprising the \textsf{B} and \textsf{C} nodes can be deleted. Besides, the guard $x > 0$ is impossible to satisfy because \textsf{A-B} is a committed state. Indeed, when the TA is in \textsf{A-B} it must transition before $x$ is incremented from zero. Hence, this transition can also be removed from the TA.
The TCTL query $\E\evt \mathit{TA}'_{tb3}.\mathsf{Error}$ checks whether or not the model verifies the mutant. If the check succeeds there is at least one execution path where $P$ and $Q$ hold at the same time.

In the RCS case study the predicates \textsf{trains\_in\_system == N} ($P$) and \textsf{trains\_in\_system == 0} ($Q$) cannot hold at the same time as trains need some time to cross the bridge.



\paragraph{Mutant 4.}
Does $Q$ ever hold? Formally, this mutant is specified as follows in MTL:  $\Box (P \rightarrow \neg \evt_{[0,\infty)} Q)$ (or as $\Box (P \rightarrow  \Box_{[0,\infty)} \neg Q)$). That is, the consequent is negated and the time interval becomes infinite.

In order to implement this mutant we first need to represent $\infty$ in Uppaal. One way of doing it is by taking $\infty$ as $k_{max}$, the maximum integer available in Uppaal. Secondly, the TA of Pattern \ref{patt:tbr} is instantiated as in Figure \ref{mut34}(\subref{fig:mut-4-TBR}), called $\mathit{TA}'_{tb4}$.
Then, $\A\Box \neg\mathit{TA}'_{tb4}.\mathsf{Error}$ is run to check whether or not the model verifies the mutant.
If the query is satisfied, $\mathit{TA}'_{tb4}$ never gets to \textsf{Error} which in turn means that $Q$ never holds in a `finite' time. Conversely, if the mutant gets to \textsf{Error} we know that $Q$ eventually holds although we do not know when it does. In any case we have gathered valuable information to fix the model. The TA of Figure \ref{mut34}(\subref{fig:mut-4-TBR}) is obtained by first mutating the original TA and then simplifying the resulting TA, as we did with \emph{Mutant 3}---i.e. Figures \ref{mut34}(\subref{fig:mut-3-TBR}-\subref{fig:mut-3-TBR-simp}).

The model of the RCS does not verify this mutant. This means that eventually all 6 trains leave the crossing area.

\subsection{\label{errorsImp}Finding timing errors in implementations}
\label{Generacion de casos de prueba}

Once the model verifies all the intended properties we can use mutants of QTP and model checking to generate test cases to test the implementation of the model\endnote{By implementation we mean the source code of an imperative (perhaps object-oriented) program implementing the model. Although in some contexts an implementation is just a more concrete model, imperative programs tend to lose many logical properties when compared with models described in terms of higher-level languages such as logic.}.

The idea is that mutants represent possible misinterpretations made by the developers. In other words, we assume developers have misunderstood the model what led them to implement a wrong model ending up in a faulty implementation. More formally, let $M$ be the model, $P$ a QTP such that $M \models P$ and $P'$ a mutant of $P$. If developers have produced a faulty implementation we can think it verifies the mutant $P'$, instead of $P$. Hence, we need traces of $M$ that do not verify $P'$---note that these traces will necessarily verify $P$ as we assume $M \models P$. Such traces can be generated by executing the following TCTL query over $M$ and the TA of $P'$: $\E\evt \TA_{P'}.error$, for some $error$ state in $\TA_{P'}$.
Actually, more such traces can be generated by considering all the  $error$ states present in $\TA_{P'}$. These traces will test the implementation in slightly different ways. 
Assume we can transform such a trace into a test case, $t$, for the implementation. If the implementation actually implements $P'$, $t$ will lead it to a state not satisfying $P$ but $P'$. Conversely, if the implementation (correctly) implements $P$, $t$ will lead it to a state satisfying $P$ and not $P'$.
Summing up, given a QTP, instantiated from one of the patterns introduced in Section \ref{Implementación de los Patrones de propiedades temporales cuantitativas}, we systematically go through all its mutants to generate counterexamples which will be transformed into test cases.

\paragraph{Useless mutants.}
Finding traces of $M$ not satisfying a mutant $P'$ is not always possible as $M$ could verify both $P$ and $P'$. For example, if the model verifies the property ``all trains leave the crossing area in at most 122 seconds'' it will also verify the mutant ``all trains leave the crossing area in at most 300 seconds''. Therefore, concerning test case generation the more interesting mutants are those that verify $M \models P \land \lnot P'$. The good thing is that if $M$ verifies $P'$ the TCTL query given above will not return a witness trace.

\paragraph{From traces to test cases.}
Once the model checker returns a trace it has to be transformed into a test case for the implementation. Below we explain how this transformation can be carried out.

In order to test a real-time system one must consider when the system has to be stimulated, when the responses should arrive and what the verdict is (an error has been found or not). Then, in the context of real-time systems a test case is a sequence of actions interleaved with delays. These actions stimulate the system after each delay. These sequences can be called \emph{timed words} \cite{AlurDill94}.
Let $A$ be a TA as defined in Section \ref{Timed Automata} and let $\Sigma$ be the set of actions of $A$. A sequence $\sigma \in (\Sigma \cup \mathbb{R}_{0}^+)^*$ is a timed word if it is of the form $\sigma= d_1\cdot a_1\cdot d_2\cdot a_2 \dots d_k \cdot a_k\cdot d_{k+1}$, where $d_i \in \mathbb{R}_0^+$ is the elapsed time between actions $a_{i-1}$ and $a_i$. 

Besides, a trace returned by a model checker is an execution of a TA. An execution of $A$ is a sequence of TA states of the form: $(n_0,u_0) \xrightarrow{\gamma_1} (n_1,u_1) \xrightarrow{\gamma_2} (n_2,u_2) \dots \xrightarrow{\gamma_n} (n_n,u_n)$, where $n_i$ is the current node of $A$, $u_i$ is the clock valuation in $n_i$ and $\gamma_j$ is either the execution of an action $\xrightarrow{g_j,a_j,r_j}$ or a delay $\xrightarrow{d_j}$ (Section \ref{Timed Automata}). Executions can be easily generalized to a network of $m$ TA executing concurrently. In effect, the state of the network is of the form $((n_i^1,\dots,n_i^m),u_i)$ where $n_i^j$ is the current node of the $j$-th TA and $u_i$ is a valuation for all the clocks of all the TA. A transition of the network is either a delay, where all the TA remain in their current nodes and the invariants are still true, or the execution of an action in one of the TA.

Hence, in order to go from traces to test cases it is necessary to transform executions of networks of TA into timed words. An execution can be transformed into a timed word by considering the following cases:
\begin{enumerate}
\item If $\gamma_i$ is $\xrightarrow{d_i}$ and $\gamma_{i+1}$ is  $\xrightarrow{g_{i+1},a_{i+1},r_{i+1}}$, then $d_i\cdot a_{i+1}$ is added to the timed word.
\item If $\gamma_i$ is $\xrightarrow{g_i,a_i,r_i}$ and $\gamma_{i+1}$ is $\xrightarrow{g_{i+1},a_{i+1},r_{i+1}}$, then $a_i\cdot 0\cdot a_{i+1}$ is added to the timed word.
\item If $\gamma_i$ is $\xrightarrow{d_i}$ and $\gamma_{i+1}$ is $\xrightarrow{d_{i+1}}$, then $d_i + d_{i+1}$ is added to the timed word.
\end{enumerate}

In general, the timed word obtained in this way will contain information not necessary for testing the implementation. Given that the network of TA includes TA representing the the system (e.g. \texttt{Talarm} and \texttt{Railroad-Controller} in Figure \ref{fig:train-alarm-rtdevs}) as well as the environment (e.g. \texttt{Train}), the timed word will contain actions of both components. However, a test case should include just the actions produced by the environment because these are the ones that will stimulate the system. Therefore, all the actions and delays produced by the TA representing the system must be removed from the timed word. The resulting timed word is the test case to be executed.

\section{Conclusions and future work}
\label{Conclusiones y Trabajos Futuros}

In this work we have presented a technique rooted in model checking but aimed at the RT-DEVS community that can be used to: formally and automatically verify an important class of recurrent quantitative temporal properties expressed as patterns of MTL formulas appearing in RT-DEVS models; find timing errors in RT-DEVS models by using mutants of those patterns; and generate test cases to test timing requirements in the implementation of those RT-DEVS models. All these activities can be performed thanks to some of the advanced features present in the Uppaal model checker.
The case study presented in this work not only exemplifies the practical application of our verification technique but also highlights its effectiveness in identifying and addressing complex quantitative temporal requirements in real-time systems.

The use of model checkers for the verification of industrial-strength systems may pose some concerns about the applicability of the technique presented in this paper. In particular the so-called state explosion problem may render our technique impractical for some real-life problems. Nonetheless, Uppaal employs several optimizations, such as model reduction techniques and zone-based abstraction \cite{Bouyer22,Behrmann02}, in order to reduce the complexity of models and improve the efficiency of the verification algorithm. These techniques work in practice as shown by several projects on the application of Uppaal to real-time, industrial-grade problems \cite{GearController98,Bowman98,WSB-Uppaal, Bakhshi21}.
Besides the case study shown in this paper, we have validated our method by applying the \emph{Time-Bounded Response} pattern to the verification of two QTP of the Gear Control System described by Lindahl et al.\cite{GearController98}. The experimental data can be found in our GitHub repository \cite{github}.
The fact that several industrial-grade systems have been verified with Uppaal provides evidence that our method could scale up to harder problems.

For now this technique has to be conducted manually. As future work we envision a software tool implementing this technique, from translating the RT-DEVS models into TA, to selecting patterns of temporal formulas, to their instantation, to running the TCTL queries, etc.
On the one hand, the translation can be implemented with model transformation tools (cf. Model-Driven Development) by using programming languages such as QVT\cite{QVT11} or ATL\cite{ATL}. On the other hand, Uppaal provides a Java API that can be used to interact with the tool in a transparent way. In this way, it is possible to develop a comprehensive tool encompassing all the steps of our technique. Moreover, it would be possible to develop a tool  transforming the very RT-DEVS models into executable code such as PowerDEVS\cite{BK11}. This approach would facilitate the integration of the technique presented in this paper with the traditional M\&S approach, thus leveraging both of them. González et al. have already applied some of these ideas to DEVS models \cite{Gonzalez15,Gonzalez16}.

It is also our intention to work on other patterns of temporal formulas such as \emph{Time-Bounded Frequency with time-out}, \emph{Security/Absence}, \emph{Time-Bounded Stability Frequency} and their corresponding mutants. In this way properties not yet supported by our technique could also be verified.

\theendnotes

\bibliography{paper} 
\bibliographystyle{SageV}

\begin{dci}
The authors declare that they have no known competing financial interests or personal relationships that could have appeared to influence the work reported in this paper.
\end{dci}

\appendix

\section{Mathematical definition of the RT-DEVS model of the RCS}
\label{RCS RT-DEVS}

This section presents the mathematical version of the RT-DEVS models Train and Alarm depicted in Figure \ref{fig:train-alarm-rtdevs}. In order to do that we consider the following simplified version of RT-DEVS:

RT-DEVS Train = $\langle X,Y,S,\delta_{ext},\delta_{int},\lambda,ti\rangle$

$X = \{stop, go\} \times \nat$

$Y = ( \{appr, leave\} \times \nat) \cup ( \{alarm\} \times \{\tau\} ) \cup \phi$ \\
where $\tau$ represents an alarm signal and $\phi$ a ``dummy'' output. Given that RT-DEVS asks for an output every time an internal transition is triggered, we use $\phi$ when this output is not expected by other RT-DEVS component.

$S = \{Safe, Talarm, Appr, Stop, Start, Cross\}$. 

$
\delta_{ext} ((s, e), (p, v)) =
\begin{cases}
    Stop, & \text{if } s=Appr \wedge p=stop\\
    & \wedge \; e \leq 10 \\
    Start, & \text{if } s=Stop \wedge p=go\\
\end{cases}
$


$
\delta_{int} (s) =
\begin{cases}
    Talarm, & \text{if } s=Safe \wedge input\_enabled \\
    Appr, & \text{if } s=Talarm \\
    Cross, & \text{if } s=Appr \vee s=Start\\
    Safe, & \text{if } s=Cross \\
\end{cases}
$



$
\lambda (s) =
\begin{cases}
    (appr, id ), & \text{if } s=Safe \\
    (alarm, \tau ), & \text{if } s=Talarm \\
    (leave, id ), & \text{if } s=Cross \\    
    \phi, &  \text{Otherwise} \\

\end{cases}
$

where $id$ is a train identifier.

$
ti (s) =
\begin{cases}
    [0,\infty] & \text{if } s=Safe \\
    [0,0], & \text{if } s=Talarm \\
    [10,20], & \text{if } s=Appr \\
    [\infty,\infty], & \text{if } s=Stop \\    
    [7,15], & \text{if } s=Start\\    
    [3,5], & \text{if } s=Cross \\    
\end{cases}
$\\

The interval $[\infty,\infty]$ represents passive states, i.e., the system can leave these states only when an input is received.


RT-DEVS Alarm = $\langle X,Y,S,\delta_{ext},\delta_{int},\lambda,ti\rangle$

$X = \{alarm\} \times \{\tau\}$.

$Y = ( \{alarm\_on, alarm\_off\} \times \{\tau\} ) \cup \phi$

$S = \{Off, Warning, Danger, Howl, On\_Off\}$. 

$ \delta_{ext} ((s, e), (p, v)) = Warning,  \text{if } s=Off \wedge p=alarm\wedge trains\_in\_system\geq 3 $

$
\delta_{int} (s) =
\begin{cases}
    Danger, & \text{if } s=Warning \wedge e\geq 2\\

    Off, & \text{if } s=Warning \wedge \\
    & trains\_in\_system\leq 3\\

    Howl, & \text{if } s=Danger \\

    On\_Off, & \text{if } s=Howl \wedge (e==3 \; \vee\\
    & trains\_in\_system<3)\\

    Off, & \text{if } s=On\_Off \wedge \\
    & !decompressing\_area \\

\end{cases}
$

$
\lambda (s) =
\begin{cases}
    (alarm\_on, \tau ), & \text{if } s=Warning \wedge e\geq 2 \\
    
    (alarm\_off, \tau ), & \text{if } s=On\_Off\\

   \phi, & \text{Otherwise}\\    
\end{cases}
$

$
ti (s) =
\begin{cases}
    [\infty,\infty] & \text{if } s=Off \\
    [0,7], & \text{if } s=Warning \\
    [7,7], & \text{if } s=Danger \\
    [0,3], & \text{if } s=Howl \\    
    [\infty,\infty], & \text{if } s=On\_Off\\    
\end{cases}
$

$input\_enabled$ and $decompressing\_area$ are Boolean variables shared between both models. $input\_enabled$ models the traffic light;  $decompressing\_area$ indicates whether the crossing are is being emptied; $trains\_in\_system$ represents the number of trains in the crossing area.

\section{More QTP Patterns}
\label{Patrones de propiedades temporales cuantitativas en Uppaal}

In this section we introduce more patterns of QTP.

\subsection{Precedence with Delay} 
\label{Precedence with Delay}

This pattern deals with situations where a predicate $P$ allows another predicate $Q$ to hold but only after some time. Then, if $Q$ holds it is because some time ago $P$ has held. As an example, a home security system becomes armed ($Q$) 10 seconds after the user has entered the security code ($P$) to allow them to leave the house.

\begin{figure}[]
   \begin{center}
     \begin{tabular}{ l }
		\begin{tikzpicture} 
			{\draw [very thick,->] (-0.8,0) -- (4,0); \filldraw [black] (-0.8,0) circle (2pt); \node (noQ) at (-0.8,0.2) {\footnotesize $\neg Q$...}
			;}
		\end{tikzpicture}  
\\                   
		\begin{tikzpicture} 
			{\draw [very thick,->] (-0.8,0) -- (5.4,0); \filldraw [black] (-0.8,0) circle (2pt); \node (noQ) at (-0.8,0.2) {\footnotesize $\neg Q$...}; \filldraw [black] (0,0) circle (2pt);\node (primerP) at (0,0.2) {\footnotesize $P$}; \filldraw [black] (4,0) circle (2pt); \node (primerQ) at (4,0.2) {\footnotesize $Q$}; \filldraw [black] (4.7,0) circle (2pt); \node (segQ) at (4.7,0.2) {\footnotesize $Q$}; \node  (intervalo) at (1.75,-0.26) {\footnotesize  k}; \draw [thick, |-|,gray] (0,-0.4) -- (3.5,-0.4);}
		\end{tikzpicture}  
  \end{tabular} 	
   \end{center}
   \caption{Traces of the \emph{Precedence with Delay} pattern not reaching \textsf{Error}}
   \label{tikz:Trazas_PrecedendeDelay} 
\end{figure}
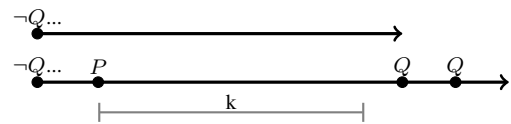

The pattern is documented in Pattern \ref{patt:pd} and some traces not reaching the \textsf{Error} state are shown in Figure \ref{tikz:Trazas_PrecedendeDelay}.

\begin{Pattern*}
\centerline{
\noindent
\fbox{%
\begin{tabularx}{.95\textwidth}{lX}
\textsc{Statement} & $P$ enables $Q$ after $k$ time-units\\
\textsc{Description} & $Q$ can hold only if $P$ holds and the current time is greater than or equal to $t(P)+k$; in other words, $P$ allows $Q$ to hold after $k$ t.u. Then, if $Q$ holds before $k$ t.u. since $P$ started to be true, the property is invalid. \\
\textsc{MTL formula} & $\evt Q \rightarrow (P \rightarrow (\square_{[0,k]} \neg Q))  \cup  Q$  \\
\textsc{Observer TA} &
	\begin{center}
		\includegraphics [scale=0.45]{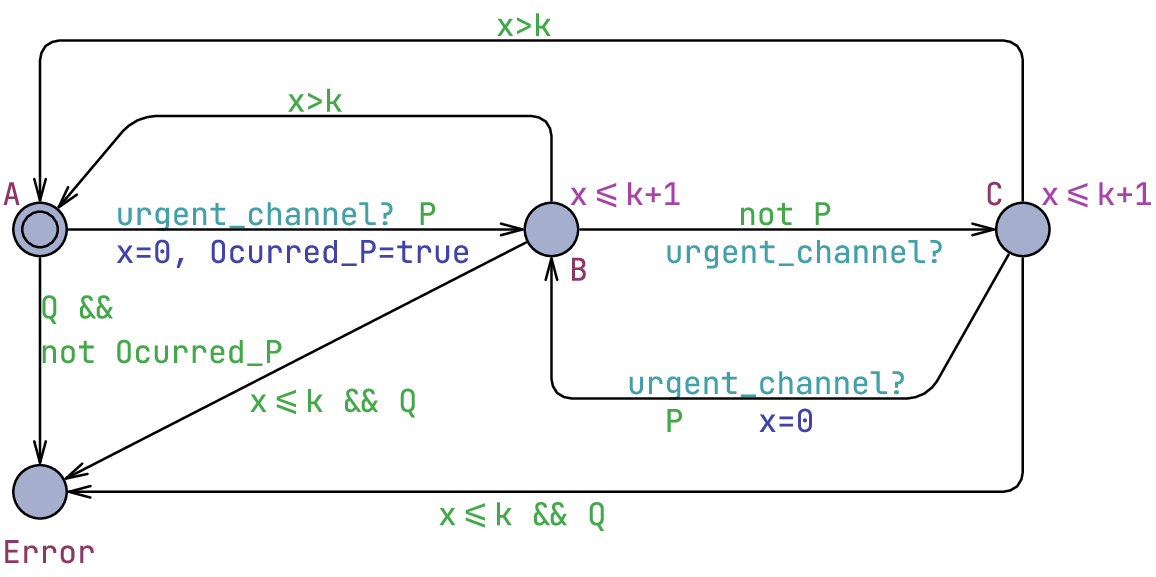}
	\end{center}
\end{tabularx}
}}
\caption{\label{patt:pd}\emph{Precedence with Delay}}
\end{Pattern*}

\subsection{Time-Bounded Frequency}
\label{Time-Bounded Frequency}

This pattern captures situations where something must hold again in the future but not too late. For instance, a data backup must be done again in at most 30 days.

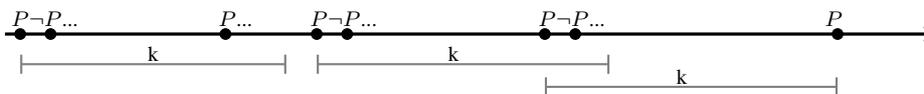
\begin{figure*} [h!]
   \begin{center}
     \begin{tabular}{ l   }                   
		\begin{tikzpicture} 
			{\draw [very thick,->] (-0.2,0) -- (12,0); \filldraw [black] (0,0) circle (2pt); \filldraw [black] (0.4,0) circle (2pt); \node (somenode) at (0.32,0.2) {\footnotesize $P  \neg P$...}; \filldraw [black] (2.7,0) circle (2pt); \node (segP) at (2.83,0.2) {\footnotesize $P$...}; \filldraw [black] (3.9,0) circle (2pt); \filldraw [black] (4.3,0) circle (2pt); \node at (4.25,0.2) {\footnotesize $P  \neg P$...};  \node  (intervalo) at (1.75,-0.26) {\footnotesize  k}; \draw [thick, |-|,gray] (0,-0.4) -- (3.5,-0.4);  \node  (intervalo) at (5.7,-0.26) {\footnotesize  k}; \draw [thick, |-|,gray] (3.9,-0.4) -- (4.25+3.5,-0.4); 
	\filldraw [black] (3.9+3,0) circle (2pt); \filldraw [black] (4.3+3,0) circle (2pt); \node at (4.25+3,0.2) {\footnotesize $P  \neg P$...};
	 \node  (intervalo) at (5.7+3,-0.56) {\footnotesize  k}; \draw [thick, |-|,gray] (3.9+3,-0.4-0.3) -- (4.25+3.5+3,-0.4-0.3); 
	 \filldraw [black] (4.3+3+3.45,0) circle (2pt); \node at (4.25+3+3.45,0.2) {\footnotesize $P$};
	 }
		\end{tikzpicture}  
	
  \end{tabular} 	

   \end{center}
   \caption{Traces of the \emph{Time-Bounded Frequency} pattern not reaching \textsf{Error}}
   \label{tikz:Trazas_TBFrequency} 
\end{figure*}

The pattern is documented in Pattern \ref{patt:pbf} and some traces not reaching the \textsf{Error} state are shown in Figure \ref{tikz:Trazas_TBFrequency}.

\begin{Pattern*}
\centerline{
\noindent
\fbox{%
\begin{tabularx}{.95\textwidth}{lX}
\textsc{Statement} & $P$ occurs frequently before $k$ t.u. \\
\textsc{Description} & Predicate $P$ holds again before $t(P)+k$ t.u.  \\
\textsc{MTL formula} & $\square \; \lozenge _{[0,k]} P$ \\
\textsc{Observer TA} &
	\begin{center}
		\includegraphics [scale=0.4]{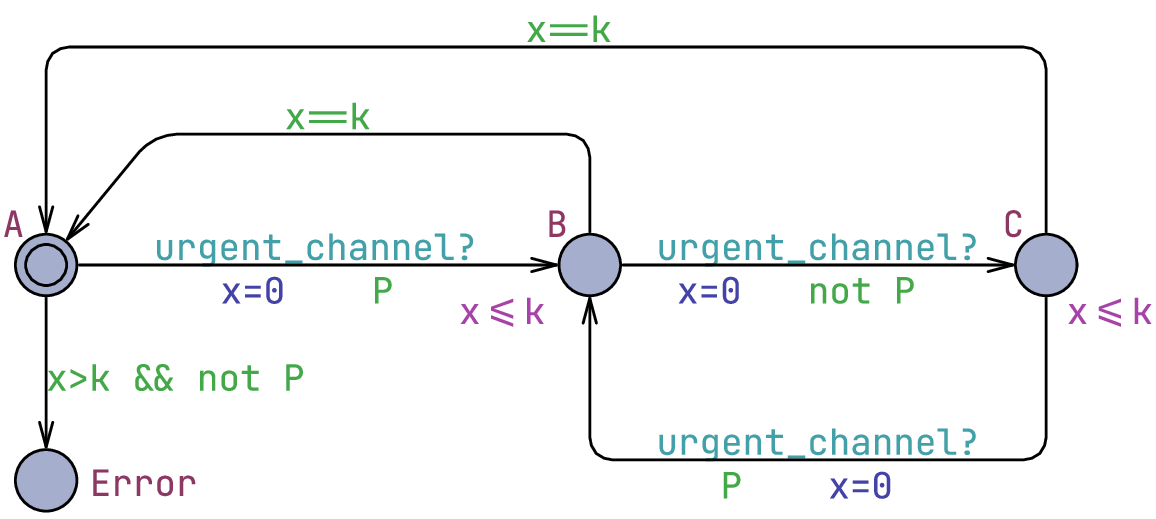}
	\end{center}
\end{tabularx}
}}
\caption{\label{patt:pbf}\emph{Time-Bounded Frequency}}
\end{Pattern*}

\subsection{Time-Constant Frequency}
\label{Time-Constant Frequency}
Some property holds periodically with a constant period. For example, a sensor must be read every 100 milliseconds. 

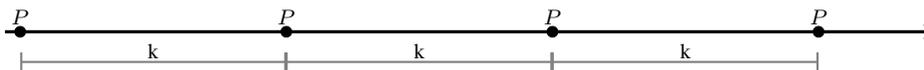
\begin{figure*} [h!]
   \begin{center}
     \begin{tabular}{ l   }                   
		\begin{tikzpicture} 
			{\draw [very thick,->] (-0.2,0) -- (12,0); 
			\filldraw [black] (0,0) circle (2pt); \node (primerP) at (0,0.2) {\footnotesize $P$};
			\node  (intervalo) at (1.75,-0.26) {\footnotesize  k}; \draw [thick, |-|,gray] (0,-0.4) -- (3.5,-0.4); 
			\filldraw [black] (3.5,0) circle (2pt); \node (segP) at (3.5,0.2) {\footnotesize $P$};
			\node  (intervalo) at (1.75+3.5,-0.26) {\footnotesize  k}; \draw [thick, |-|,gray] (3.49,-0.4) -- (3.5+3.5,-0.4); 
			\filldraw [black] (3.5+3.5,0) circle (2pt); \node at (3.5+3.5,0.2) {\footnotesize $P$};
			\node  (intervalo) at (1.75+3.5+3.5,-0.26) {\footnotesize  k}; \draw [thick, |-|,gray] (3.5+3.49,-0.4) -- (3.5+3.5+3.5,-0.4); 
			\filldraw [black] (3.5+3.5+3.5,0) circle (2pt); \node at (3.5+3.5+3.5,0.2) {\footnotesize $P$};

			}
		\end{tikzpicture}  
	
  \end{tabular} 	

   \end{center}
   \caption{Traces of the \emph{Time-Constant Frequency} pattern not reaching \textsf{Error}}
   \label{tikz:Trazas_TCFrequency} 
\end{figure*}

The pattern is documented in Pattern \ref{patt:pcf} and some traces not reaching the \textsf{Error} state are shown in Figure \ref{tikz:Trazas_TCFrequency}.

\begin{Pattern*}[ht]
\centerline{
\noindent
\fbox{%
\begin{tabularx}{.95\textwidth}{lX}
\textsc{Statement} & $P$ occurs every $k$ time-units \\
\textsc{Description} & Property $P$ always holds in $t(P)+k$ t.u.  \\
\textsc{MTL formula} & $\square \; ((P \wedge \square _{(0,k)} \neg P) \rightarrow ( \lozenge _{(0,k]} P))$ \\
\textsc{Observer TA} &
	\begin{center}
		\includegraphics [scale=0.35]{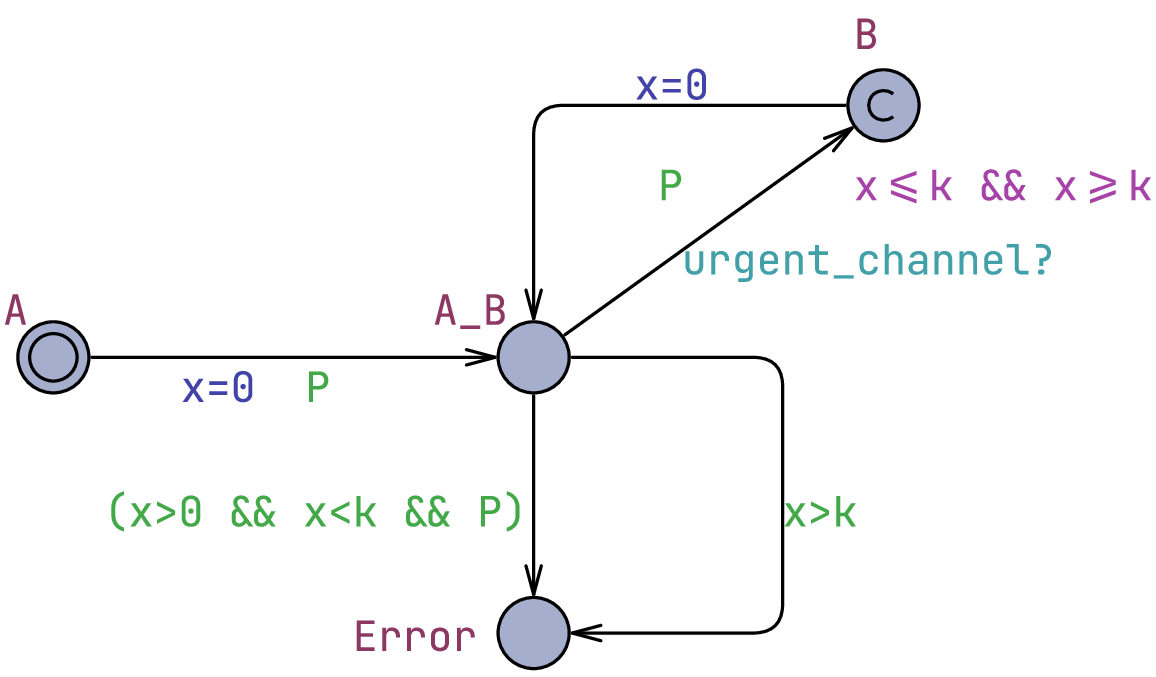}
	\end{center}
\end{tabularx}
}}
\caption{\label{patt:pcf}\emph{Time-Constant Frequency}}
\end{Pattern*}

\subsection{Time-Restricted Disable}
\label{Time-Restricted Disable}
If $P$ holds, $Q$ must hold before $k$ t.u. since $P$ became true, and when $Q$ becomes true, $P$ does not hold anymore; i.e., $Q$ deactivates $P$ in no more than $k$ t.u. Hence, if $Q$ deactivates $P$ but after $k$ t.u. or if $P$ becomes false within the $k$ t.u. interval without $Q$ being true, the property is invalid. 
For example, once a traffic light becomes green, it must turn to yellow in no more than 20 seconds.

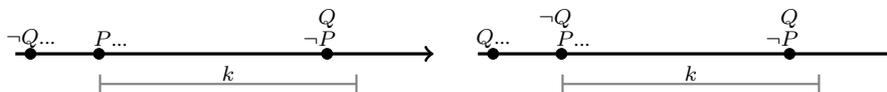
\begin{figure*}[h!]
   \begin{center}
     \begin{tabular}{ l  l }                   
		\begin{tikzpicture} 
			{\draw [very thick,->] (-1.1,0) -- (4.4,0); 
			\filldraw [black] (-0.9,0) circle (2pt); \node  at (-0.9,0.2) {\footnotesize $\neg Q$...}; 
			\filldraw [black] (0,0) circle (2pt); 	\node at (0.15,0.2) {\footnotesize $P$...}; 
			\filldraw [black] (3,0) circle (2pt); \node at (2.9,0.2) {\footnotesize $\neg P$}; \node  at (3,0.47) {\footnotesize $Q$}; 
			\node  (intervalo) at (1.7,-0.26) {\footnotesize $k$}; \draw [thick, |-|,gray] (0,-0.4) -- (3.4,-0.4); 
			}
		\end{tikzpicture}  
& 
		\begin{tikzpicture} 
			{\draw [very thick,->] (-1.1,0) -- (4.4,0); 
			\filldraw [black] (-0.9,0) circle (2pt); \node  at (-0.9,0.2) {\footnotesize $Q$...}; 
			\filldraw [black] (0,0) circle (2pt); 	\node at (0.15,0.2) {\footnotesize $P$...}; \node  at (-0.08,0.47) {\footnotesize $\neg Q$} ;
			\filldraw [black] (3,0) circle (2pt); \node at (2.9,0.2) {\footnotesize $\neg P$}; \node  at (3,0.47) {\footnotesize $Q$}; 
			\node  (intervalo) at (1.7,-0.26) {\footnotesize $k$}; \draw [thick, |-|,gray] (0,-0.4) -- (3.4,-0.4); 
			}
		\end{tikzpicture}  
  \end{tabular} 	

   \end{center}
   \caption{Traces of the \emph{Time-Restricted Disable} pattern not reaching \textsf{Error}}
   \label{tikz:Trazas_TRDisable} 
\end{figure*}

The pattern is documented in Pattern \ref{patt:ptrd} and some traces not reaching the \textsf{Error} state are shown in Figure \ref{tikz:Trazas_TRDisable}.

\begin{Pattern*}[ht]
\centerline{
\noindent
\fbox{%
\begin{tabularx}{.95\textwidth}{lX}
\textsc{Statement} & $Q$ disables $P$ within $k$ time-units \\
\textsc{Description} & When $P$ becomes true, $Q$ will be true in $t(P)+k'$ with $k'\leq k$ and in that moment $P$ will not hold anymore.  \\
\textsc{MTL formula} & $\square \; (P \rightarrow ((P \wedge \neg Q) \cup _{[0,k]} (Q \wedge \neg P)))$ \\
\textsc{Observer TA} &
	\begin{center}
		\includegraphics [scale=0.56]{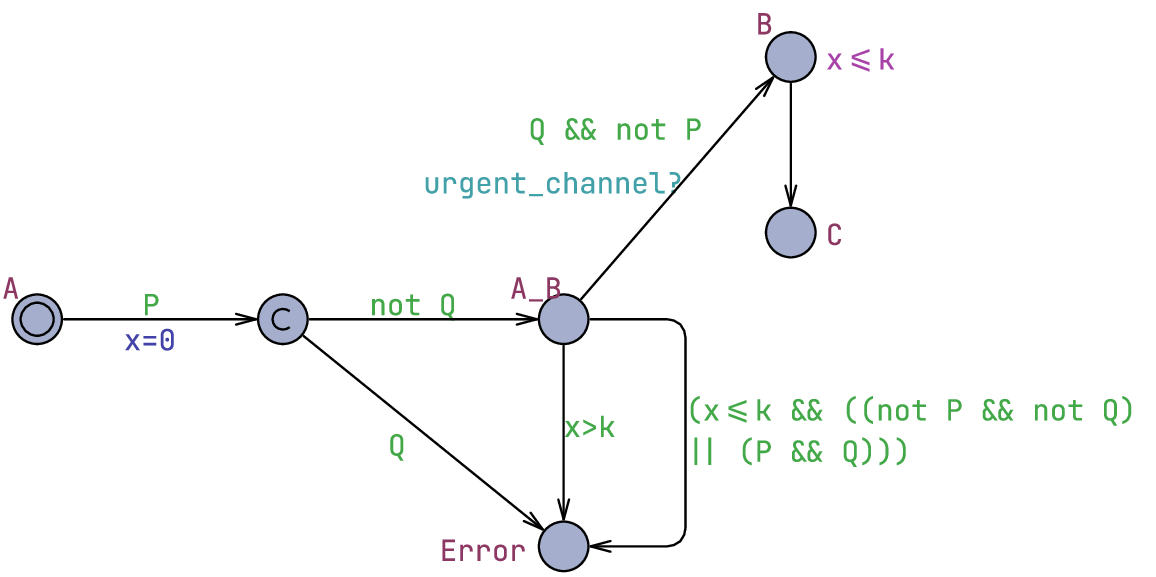}
	\end{center}
\end{tabularx}
}}
\caption{\label{patt:ptrd}\emph{Time-Restricted Disable}}
\end{Pattern*}

\end{document}